\documentclass{interact}
\usepackage{subfigure}
\usepackage{epstopdf}
\usepackage[caption=false]{subfig}

\usepackage[longnamesfirst,sort]{natbib}
\bibpunct[, ]{(}{)}{;}{a}{,}{,}

\usepackage{amsmath}
\usepackage{amsthm}

\usepackage{multirow}

\theoremstyle{plain}

\theoremstyle{definition}

\theoremstyle{remark}

\begin{document}

\title{A linear programming approach for designing multilevel PWM waveforms}

\author{Shravan Mohan and Bharath Bhikkaji\\Department of Electrical Engineering,\\ Indian Institute of Technology Madras, Chennai.}

\date{}

\maketitle

\begin{abstract}
This paper considers the problem of designing a multilevel pulse width modulated waveform (PWM) with a prescribed harmonic content. Multilevel PWM design plays a major role in many diverse engineering disciplines.  In power electronics, multilevel  PWM design corresponds to determining the  inverter switching times and levels for selective harmonic elimination and harmonic compensation. In mechatronics, the same design corresponds to shaping input signals to damp residual vibrations in flexible structures. More generally, in most applications, the aim of PWM design is to minimize the  total harmonic distortion  while adhering to a prescribed harmonic content. The solution approach presented in this paper is based on linear programming with the objective of minimizing the total harmonic distortion. This objective is achieved within an arbitrarily small bound of the optimal solution. In addition, the linear programming formulation makes the design of such switching waveforms computationally tractable and efficient. Simulations are provided for corroboration.
\end{abstract}

\section{Introduction}
\begin{figure}[t]
\centering
\includegraphics[width=4.8in]{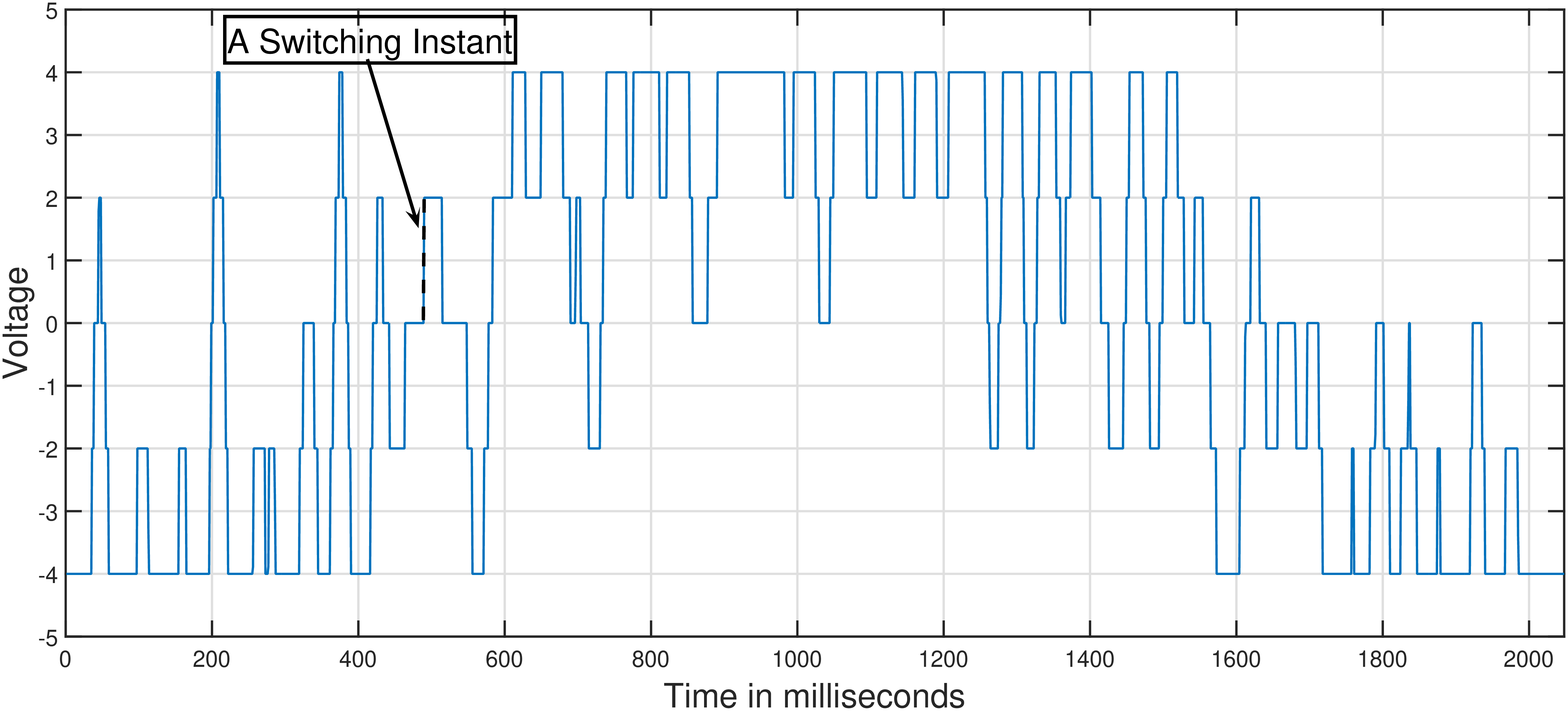}
\caption{This figure shows a multilevel PWM. The signal takes has a period of 2.048 milliseconds and takes values from the set $\{-4, ~~-2, ~~0, ~~2 ,~~4\}$. A switching instant is also shown in the figure where the signal transition from 0 to 2.}
\label{fig:waveform_example}
\end{figure}

In this paper, the problem of designing a multilevel pulse width modulated waveform (PWM) with a prescribed harmonic content is considered. A multilevel PWM is a periodic  signal which takes values from a finite discrete set of real numbers. The time instants where it changes value are called switching instants. Figure \ref{fig:waveform_example} shows one period of a typical multilevel PWM. The prescribed harmonic content specifies a finite number of Fourier components which the multilevel PWM must contain. Since the multilevel PWM is a switching signal, it also contains harmonics other than the ones prescribed. The harmonic content in the multilevel PWM which are not part of the prescribed set gives rise to  harmonic distortion. It is desired that this distortion be kept as low as possible in all applications. Therefore, the objective of the design of a multilevel PWM refers to the determination of its switching instants and the level transitions such that: (i) the signal has the prescribed harmonic content and (ii) the total harmonic distortion is minimized. 

Switched waveforms play an important role in many engineering domains such as power electronics, mechanical systems and power amplifiers. The switched waveforms form an integral part of the operation of power electronic systems, in particular of that of power inverters (\cite{holmes2003pulse},  \cite{dahidah2008selective}, \cite{sheng2016improved}). A power inverter converts DC power to AC power. The output of a power inverter is a periodic multilevel PWM (typically with a period of  20 ms). Thus, its harmonic content can be controlled by choosing appropriate switching instants and level transitions. Conventionally, if the prescribed values for the higher order harmonics are all zero (higher order refers to harmonics other than the fundamental harmonic), then the problem is called Selective Harmonic Elimination (SHE). On the other hand, if the prescribed values for the higher order harmonics have at least one non-zero number, the problem is called Harmonic Compensation (HC). In the context of mechanical systems, switched waveforms are used in damping residual vibrations. Here, the main idea is to generate a multilevel PWM consisting of an out of phase harmonic at the resonant frequency of the mechanical system. It is also desirable to design a switching signal with the lowest harmonic distortion so that effect of PWM signal on the intended operation of mechanical system is minimal. Refer to \cite{gurleyuk2011designing} and \cite{song1999spacecraft} for   examples on vibration damping using  multilevel PWM signals. Another application of multilevel PWMs is in switch-mode power amplifiers which are crucial components of modern day Radio Frequency (RF) transmitters used in wireless communication systems. The key metric for evaluating the performance of a switch-mode power amplifiers is called the power coding efficiency (see \cite{zhu20145}, \cite{chung2015outphasing}, \cite{francois2014reconfigurable}). The power coding efficiency is defined as the ratio of the prescribed in-band power to the total band power of the digitized signal. A high power coding efficiency can lead to an economical filter design at the receiver. In addition to these applications, multilevel PWMs are also used in sigma-delta modulation (\cite{i2013study}) and battery management systems (\cite{maharjan2012active}).   

The problem of designing multilevel PWMs has been widely studied in the context of SHE and HC and various solution methodologies have been proposed. The conventional method is to write the harmonic quantities as trigonometric polynomials of the switching angles and find a solution to these equations\footnote{For example, in the case of cascaded inverters, the authors of \cite{dahidah2008selective} show that, with the assumption of half wave symmetry, the $k^{\rm{th}}$ sine Fourier component is given by: $\displaystyle b_k = \frac{4V_{dc}}{k\pi} \left(V_1\sum_{i=1}^{P_1}(-1)^{i+1}\cos(k\alpha_i) ~+~\dots~ + V_M(-1)^{P_{M-1}+1}\sum_{i=P_{M-1}}^{P_M}(-1)^{i+1}\cos(k\alpha_i)\right)$, where $M$ is the number of levels, $V_{dc}$ is the DC bus voltage, $V_{dc}V_j$ is the value of the $j^{\rm{th}}$ level and $\alpha_i$ is the $i^{\rm{th}}$ switching angle. And if $N_j$ is the number of quarter wave switches at the $j^{\rm{th}}$ level converter, $P_j = N_1 + \dots +N_j$.}. In many of the previous works, these polynomial equations are used as constraints in an optimization problem where the cost function is the total harmonic distortion.   These optimization problems are generally non-convex and hence, do not have efficient solvers. Among the many techniques employed for solving such optimization problems are variants of Genetic Algorithms, Simulated Annealing, Artificial Neutral Networks, Particle Swarm Optimization, Bee Algorithm, Fuzzy Logic, Frog Leaping Algorithm and random search based heuristics. The reader is referred to  \cite{dahidah2008selective},  \cite{haghdar2011selective}, \cite{lee2015switched}, \cite{maia2013adaptive},  \cite{kumle2015application}, \cite{kavousi2012application},   \cite{lou2014fundamental} \cite{sheng2016improved}, \cite{fisher1994continuous}, \cite{kumar2005bi}, \cite{lohia2008minimally}, \cite{nanda2006novel}, \cite{qianimproved} and \cite{franquelo2007flexible} for more details. Another broad set of methods depend on gradient search, see for example \cite{agelidis2008attaining}. Algebraic techniques have also received attention as a solution methodology, for example, methods which use resultants of polynomials, Grobner basis, Chebyshev functions and Walsh functions. For a detailed exposition on these methods, see  \cite{yang2016selective},  \cite{pindado1998robust} and \cite{liang1997inverter}. Space Vector Modulation techniques provide another route for designing multilevel PWMs. These methods are computationally efficient, but these need not produce waveforms with low THD. Refer to \cite{malinowski2010survey} and \cite{mcgrath2003optimized}  for a detailed insight on Space Vector Modulation methods for multilevel inverter switching.   

The proposed algorithm of this paper has the following advantages over the above references:
 \begin{itemize}
\item The proposed algorithm relies on linear programming which is computationally efficient.
 \item The solution can be made to satisfy the harmonic constraints within an arbitrarily small bound and its THD to lie within  an arbitrarily small bound of the optimal value.
 \item The algorithm does not require an initial guess of the solution. A method like the simplex calculates an initial feasible solution \cite{bertsimas1997introduction}.
 \end{itemize}

The organization of the paper is as follows. The mathematical problem and the details of the solution are presented in Section 2. Simulations for various scenarios of SHE and HC are presented in Section 3. The designed multilevel PWMs, their harmonics and the total harmonic distortion are shown in figures and tables. Conclusions and future directions of work in this regard are outlined in last section which is followed by references.

\section{Problem description and the proposed solution}
The following are some of the mathematical notations used in this paper; see~\cite{bernstein2009matrix} for more details.
\begin{itemize}
\item $\displaystyle \mathcal{R}$ and $\displaystyle \mathcal{C}$ denote the \textit{real} and \textit{complex} number fields, respectively. The set of positive integers is denoted by $\mathcal{Z}$.
\item For a set $S$, $\displaystyle |S|$ denotes its \textit{cardinality}.
\item For a complex number $x$, $|x|$ is its \textit{absolute value}.
\item For a vector $x\in \mathcal{C}^{n\times 1}$, $\Re(x)$ and $\Im(x)$ are vectors of its real and imaginary components, respectively.
\item $\displaystyle \textbf{1}^{m\times n}$ denotes a matrix of ones of dimension $\displaystyle (m\times n)$.
\item For a matrix $A$, $A(i,:)$ denotes its $i^{\rm{th}}$ row and $A(:,j)$ its $j^{\rm{th}}$ column, $A^T$  its \textit{transpose}, $\displaystyle \mbox{dim}(A)$ gives the dimension of its column space 	 (also called its \textit{rank}) and $\displaystyle \mbox{vec}(A)$ denotes the column-wise  \textit{vectorization} of $A$, {\it i.e.,} $\displaystyle \mbox{vec}(A) = \left[A(:,1)^T~ A(:,2)^T~\dots~A(:,n)^T\right]^T$ .
\item For matrices $A$ and $B$, $\displaystyle A \otimes B$ denotes their \textit{Kronecker product}.
\item For a vector $x$, $\displaystyle ||.||_{k}$ is the $\displaystyle k-$\textit{norm}.
\item A matrix $A$ is called a \textit{stochastic matrix} if each of its elements of is non-negative and each row adds to 1. 
\item $\{0,1\}^{N\times S}$ denotes the set of all  $N \times S$ matrices with elements being either $0$ or $1$. 
\end{itemize}

\subsection{\textit{Problem description}}
A multilevel PWM is a periodic signal $x(t)$, with period $T$, which takes values from a finite subset of $\mathcal{R}$. Let $\displaystyle L = \{L_1,\dots, L_m\}$ be that subset of $\mathcal{R}$. The multilevel PWM must also consist of a finite set of prescribed harmonic content. Suppose that the prescribed harmonic numbers, and their values, are given by the set $K = \{k_1, \dots, k_r\}$ and the set $\displaystyle H = \{h^c_{k_1} + jh^s_{k_1}, \dots, h^c_{k_r}+jh^s_{k_r}\}$, respectively. For ease of exposition, denote the vectors $h = \left[h^c_{k_1} + jh^s_{k_1}~ \dots ~h^c_{k_r}+jh^s_{k_r}\right]^T$, $h_c = \left[h^c_{k_1}~ \dots ~h^c_{k_r}\right]^T$ and $h_s = \left[h^s_{k_1}~ \dots ~h^s_{k_r}\right]^T$. Under the aforementioned constraints, the objective is to design a multilevel PWM with the lowest total harmonic distortion (THD), which is defined as
\begin{eqnarray}
\nonumber {\rm THD}(x) &=& \frac{\displaystyle\sum_{k\notin K}|F_{k}(x(t))|^2}{\displaystyle\sum_{k\in \mathcal{Z}}|F_{k}^Tx|^2} 
\\ &=& 1 - \frac{|F_{k_1}(x(t))|^2 + |F_{k_2}(x(t))|^2 + \dots + |F_{k_r}(x(t))|^2}{|F_1(x(t))|^2+|F_2(x(t))|^2 + |F_3(x(t))|^2 + \dots},
\label{eqn:THD_definition}
\end{eqnarray}
where
\begin{eqnarray}
\nonumber F_k(x(t)) = \displaystyle \frac{2}{T}\int_{t=0}^{T} x(t)\exp\left(-j\frac{2\pi kt}{T}\right)dt
\end{eqnarray}
are the fourier coefficients.
Putting it all together, the problem in hand can be stated as: 
\begin{equation}
\begin{array}{l}
\displaystyle \mbox{find}~~ x(t),~ t \in [0, T), \mbox{~with the lowest THD} \\
\displaystyle \mbox{such that}~\left\{ \begin{array}{l}
  \displaystyle  x(t) \in L, \forall t \in [0, T)\\
  \displaystyle  \frac{2}{T}\int_0^T x(t)\exp{\left(-j\frac{2\pi kt}{T}\right)} dt = h^c_k + jh^s_k, 
  \forall k \in \{k_1, \dots, k_r\}.
       \end{array} \right.
\end{array} 
\label{eqn:prob_definition}
\end{equation}
In addition, the following assumptions are made:
\begin{enumerate}
\item[i.] The value of the fundamental harmonic is part of the prescribed set and $h$ is assumed to be a non-zero vector.
\item[ii. ] The set $L$ consists of at least 3 real numbers and $L_1 <  L_2 \dots L_{m-1} < L_m$. A linear programming approach to design waveforms for two-level inverters has been presented in \cite{mohan2017linear}.
\item[iii.] The signal $\displaystyle x(t)$ is piece-wise constant within the time intervals demarcated by $\displaystyle t_i = (i-1)T/N$, $\displaystyle i\in {1, \dots, N}$. This implies that the switching instants are restricted to $\displaystyle t_i$'s and thus the optimization problem \eqref{eqn:prob_definition} can be cast over the set of discrete variables given by  $\displaystyle \left\{x\left(t_i\right)| i\in \mathcal{Z} \mbox{~and~} 1\leq i\leq N\right\}$. Suppose $\displaystyle x\left(t_i\right)$ is written as $\displaystyle x(i)$. With a slight abuse of notation, let $x = \left[x(1)~x(2)~ \dots~x(N)\right]^T$.
\item[iv.] {$N$ is chosen such that $\displaystyle r\ll N$ and $\max(K)\ll N$, where $r=|K|$. This assumption allows for the $\displaystyle k^{\rm{th}}$ ($k\in K$) Fourier component to be approximated as:
\begin{equation}
\begin{array}{l}
F_k(x(t)) = \displaystyle \frac{2}{T}\int_{t=0}^{T} x(t)\exp\left(-j\frac{2\pi kt}{T}\right) dt \approx \frac{2}{N}\sum_{i=0}^{N-1}x\left(i\right)\exp\left(-j\frac{2\pi k i}{N}\right) = f_k^T x.
\end{array}
\label{eqn:fourierapprox}
\end{equation}}
\end{enumerate}

From the assumptions iii.  and iv. the definition of THD can be recast as:
\begin{equation}
\displaystyle \mbox{THD}(x) = \frac{\displaystyle\sum_{k\notin K}|f_{k}^Tx|^2}{\displaystyle\sum_{k\in \mathcal{Z}}|f_{k}^Tx|^2} = 1 - \frac{\displaystyle\sum_{k\in K}|f_{k}^Tx|^2}{\displaystyle\sum_{k\in \mathcal{Z}}|f_{k}^Tx|^2}.
\label{eqn:THD_definition_2}
\end{equation}
Similarly the optimization problem \eqref{eqn:prob_definition} can also be rewritten as
\begin{equation}
\begin{array}{l}
\displaystyle \min_{x\in \mathcal{R}^N} ~~~~\mbox{THD}(x) \\
\displaystyle \mbox{subject to}~\left\{ \begin{array}{l}
  \displaystyle  x \in L^N\\
  \displaystyle  f_k^T x = h^c_k + jh^s_k, 
  \forall k \in \{k_1, \dots, k_r\}.
       \end{array} \right.
\end{array} 
\label{eqn:prob_definition_2}
\end{equation}
The optimization problem in \eqref{eqn:prob_definition_2} will be the focus of this paper in the following sections. 

\subsection{\textit{The proposed algorithm}}
The algorithm shown in Figure \ref{fig:MainLP} is the one proposed in this paper. 
The algorithm has two parts: (i) a linear program and (ii) a clamping operation on the solution obtained from the linear program. The output of the algorithm is the denotes as the vector $x$. This section discusses the constraints used in the linear program, the construction of the cost function and the effects of the clamping operation on the optimality of the final solution. It is shown that, as the time discretization of the waveform (decided by the parameter $N$) becomes finer, the solution converges to the true optimal solution, in terms of its total harmonic distortion.  
\subsubsection{The constraints}
The optimization problem in \eqref{eqn:prob_definition_2} has two constraints:\\ 
$$(i)\,x\in L^N  \,\text{~and~}\, (ii)\,f_k^Tx = h_k, ~\forall k\in K.$$ The second set of constraints which enforce the multilevel PWM to have the desired harmonics  are linear. The first constraint, which ensures that the solution is multilevel, is not a convex constraint. Hence, a natural approach is to find a convex relaxation  for it. To this end, first denote $S=\left[L_1,~\dots~,L_m\right]^T$. It is clear that
\begin{equation}
\begin{array}{l}
\displaystyle\left\{x\in L^{N\times 1}\right\} \Rightarrow \left\{\exists~ Z\in \{0,1\}^{N\times S} \mbox{~s.t.~} x=Z\times S\right\}. 
\end{array}
\label{eqn:relax1}
\end{equation}
Therefore, the optimization problem \eqref{eqn:prob_definition_2} can be written in terms of $Z\in \mathcal{R}^{N\times m}$ as:
\begin{equation}
\begin{array}{l}
\displaystyle \min_{Z\in \mathcal{R}^{N\times m}} ~~~~\mbox{THD}(x) \\
\displaystyle \mbox{subject to}~\left\{ \begin{array}{l}
x = Z\times S,\\
  \displaystyle  Z \in \{0,1\}^{N\times m}\\
  \displaystyle  f_k^T x = h(k), 
  ~~\forall k \in \{k_1, \dots, k_r\}.
       \end{array} \right.
\end{array} 
\label{eqn:prob_definition_3}
\end{equation}
Note that the convex hull of the set of matrices $\{0,1\}^{N\times m}$ such that each of the rows add to 1 is the set of stochastic matrices of dimension $N\times m$. This leads to the following relaxed optimization problem given by:
\begin{equation}
\begin{array}{l}
\displaystyle \min_{Z\in \mathcal{R}^{N\times m}} ~~~~\mbox{THD}(x) \\
\displaystyle \mbox{subject to}~\left\{ \begin{array}{l}
x = Z\times S,\\
  \displaystyle  Z \geq 0, ~~Z\times \textbf{1}^{m\times 1} = \textbf{1}^{N\times 1}\\
  \displaystyle  f_k^T x = h(k), 
  ~~\forall k \in \{k_1, \dots, k_r\}.
       \end{array} \right.
\end{array} 
\label{eqn:prob_definition_4}
\end{equation}
Note that the constraints of \eqref{eqn:prob_definition_4} are the same as the proposed algorithm in \eqref{eqn:MainLP}. The next subsection would motivate the construction of the cost function which would finally lead to the linear program in \eqref{eqn:prob_definition_4}. 
\subsubsection{The cost function}
With the constraints in place, it now remains to construct the cost function used in \eqref{eqn:MainLP}. Note that 
\begin{equation}
\displaystyle\sum_{k\in K}|f_{k}^Tx|^2 = \sum_{k\in K}|h(k)|^2,
\end{equation}
which is a constant.
Therefore, using the definition in \eqref{eqn:THD_definition_2}, minimizing THD($x$) is equivalent to minimizing $\displaystyle\sum_{k\in \mathcal{Z}}|f_{k}^Tx|^2$. By Parseval's theorem
\begin{equation}
\displaystyle\sum_{k\in Z}|f_{k}^Tx|^2 = \frac{1}{N}\sum_{i=1}^{N}x^2(i), \mbox{~~the total energy in $x$.}
\end{equation}
Suppose $\displaystyle S_p=\left[L_1^2,~\dots~,L_m^2\right]$ and $\displaystyle c = \displaystyle \frac{\textbf{1}^{1\times N}\otimes S_p}{N}$. Then note the following implication
\begin{equation}
\begin{array}{l}
\displaystyle\left\{x\in L^{N\times 1}\right\} \Rightarrow \left\{\exists~ Z\in \{0,1\}^{N\times S} \mbox{~s.t.~} x=Z\times S\right\} 
\displaystyle\Rightarrow \left\{\frac{1}{N}\sum_{i=1}^{N}x^2(i) = c^T\mbox{vec}(Z)\right\}.
\end{array}
\label{eqn:linearity_cost_function}
\end{equation}
With some algebraic manipulations, it can be seen that 
\begin{equation}
\displaystyle c^T\mbox{vec}(Z) = \left(\frac{\textbf{1}^{1\times N}\otimes S_p}{N}\mbox{vec}(Z)\right) = \frac{1}{N}\left(\textbf{1}^{1\times N}\times Z\times S_p^T\right).
\label{eqn:cost_function_1}
\end{equation}
Note that the last term in \eqref{eqn:cost_function_1} is the cost function used in \eqref{eqn:MainLP}, the linear program of the proposed algorithm. It is now imperative to inspect the quality of the solution obtained by solving the linear programming problem in \eqref{eqn:MainLP}. Moreover, it is also necessary to assess the impact of the clamping operation that follows the linear program in \eqref{eqn:MainLP}. These aspects are discussed in the subsections that follow.
\begin{figure}[t]
\centering
\fbox{\begin{minipage}{35em}
\begin{equation}
\begin{array}{l}
{\displaystyle \hspace{-1cm}\mbox{Solve}~~\min_{Z\in \mathcal{R}^{N\times m} }} ~~~\displaystyle\frac{1}{N}\left(\textbf{1}^{1\times N}\times Z\times S_p^T\right) \\\\
\displaystyle \mbox{subject to}~\left\{ \begin{array}{l}
 x = Z\times S,\\
\displaystyle   Z \geq 0, ~~Z\times \textbf{1}^{m\times 1} = \textbf{1}^{N\times 1} \\
 \displaystyle  f_k^Tx = h(k), ~~\forall k \in \{k_1, \dots,k_r\}, 
         \end{array} \right.
\end{array}
\label{eqn:MainLP}
\end{equation}
\hspace{1.8cm}where $S=\left[L_1,~\dots~,L_m\right]^T$ and $S_p=\left[L_1^2,~\dots~,L_m^2\right]$.\\\\
\hspace*{0.7cm} Clamp elements of $x$ not in $L$ to the nearest level in $L$.
\end{minipage}}
\caption{The proposed algorithm of this paper. The first step is to solve the linear program given by \eqref{eqn:MainLP}. The second, and the last, step is a clamping operation.}
\label{fig:MainLP}
\end{figure}

\subsubsection{The quality of the LP solution}
This section discusses two aspects of the quality of the solution obtained from the linear program in \eqref{eqn:MainLP}. The first objective  is to determine whether the elements of the solution vector $x$ obtained from the linear program belong to the set $L$. To that end, by substituting $x=Z\times S$ and using the fact that for matrices $U, X, V, W$ of appropriate dimensions $\{U X V^T=W\} \Leftrightarrow \{(U\otimes V)\mbox{vec}(X)=\mbox{vec}(W)\}$, the linear program in \eqref{eqn:MainLP} is written equivalently as:
\begin{equation}
\begin{array}{l}
{\displaystyle \min_{Z\in \mathcal{R}^{N\times m} }} ~\displaystyle ~~c^T\mbox{vec}(Z) \\
\displaystyle \mbox{subject to}~\left\{ \begin{array}{l}
\displaystyle  \begin{bmatrix}
          \Gamma_1   \\
           \Gamma_2
\end{bmatrix} \mbox{vec}(Z) = \begin{bmatrix}
          b  \\
           \textbf{1}
\end{bmatrix} \\\\
\displaystyle  ~Z >= 0, 
\end{array} \right.
\end{array}
\label{eqn:LP_relaxation}
\end{equation}
where
\begin{equation}
\centering
\begin{array}{l}
\hspace{1.5cm}\Gamma_1 = A\otimes S^T,~ \Gamma_2 =  \underbrace{\begin{bmatrix}I_{N\times N} \dots I_{N\times N}\end{bmatrix}}_{m},\\ ~~~~~A = \left[\Re(f_{k_1})~\dots~\Re(f_{k_r})~\Im(f_{k_1})~\dots~\Im(f_{k_r})\right]^T,\\~~b = \left[h^c_{k_1}~\dots~h^c_{k_r}~h^s_{k_1}~\dots~h^s_{k_r}\right]^T \mbox{~~and~~} c = \displaystyle \frac{\textbf{1}^{1\times N}\otimes S_p}{N}.
\end{array}
\label{eqn:ABS_definitions}
\end{equation}
Now, the fundamental theorem of linear programming aids the analysis. The fundamental theorem of linear programming states that solution (if it exists) to the standard linear program given by
\begin{equation}
\begin{array}{l}
\displaystyle \min_{x\in \mathcal{R}^N} \hspace*{1cm} p^T x;~ p\in \mathcal{R}^{N} \\
\mbox{subject to~~} \\
\hspace*{2cm}Qx = S;~~Q\in \mathcal{R}^{r\times N} ~\&~ S\in \mathcal{R}^r\\ 
\hspace*{2cm} x >= 0
\end{array}
\end{equation}
is a basic feasible solution. Moreover, when the number of constraints are lesser than the number of variables $(r<N)$, a basic feasible solution vector has at most $N-\mbox{Rank(Q)}$ non-zero elements. In \eqref{eqn:LP_relaxation}, the analogue of the matrix $Q$ is $\displaystyle \begin{bmatrix} \Gamma_1 \\ \Gamma_2 \end{bmatrix} $. Its rank can be bounded by:   
\begin{equation}
\begin{array}{l}
\mbox{Rank}\left(\begin{bmatrix}
         \Gamma_1   \\
           \Gamma_2
\end{bmatrix}\right) \leq \mbox{Rank} \left(\Gamma_2\right)+\mbox{Rank}\left(\Gamma_1\right) = N + 2r.
\end{array}
\label{eqn:rank}
\end{equation}
Suppose that the solution to the linear program in \eqref{eqn:LP_relaxation} is given by $Z^*$. Then $\displaystyle Z^*$ would have at least $\displaystyle (Nm-N-2r)$ zero elements. Suppose that the number of zero elements in the $\displaystyle k^{\rm{th}}$ row of $\displaystyle Z^*$ is given by $\displaystyle q_k$. Then
\begin{equation}
q_1 + q_2 + ~\dots~ +q_{N-1} +q_N \geq Nm- N -2r.
\label{eqn:zerosum}
\end{equation}
Since $\displaystyle Z^*$ is stochastic (each row adds to 1), 
\begin{equation}
\displaystyle q_k\leq (m-1), ~~\forall~ 1\leq k\leq N,~ k\in \mathcal{Z}.
\label{eqn:stochastic}
\end{equation}
Therefore, \eqref{eqn:zerosum} and \eqref{eqn:stochastic} together imply that at least $\displaystyle (N-2r)$ of $\displaystyle q_k$'s have to be equal to $\displaystyle (m-1)$. That is, $\displaystyle Z^*$  has at least $\displaystyle (N-2r)$ rows with $(m-1)$ zero elements. In other words, $\displaystyle x^*=Z^*\times S$ must have at least $(N-2r)$ elements in $\displaystyle L$. 

Since the aim of this paper is to find a strictly multilevel solution (each element of the solution vector must belong to the set $L$), the elements of $x^*$ not in the set $L$ need to be modified. The modification is by clamping such an element of $x^*$ to its nearest value in $L$. This operation is equivalent to quantization and is also last step mentioned in the proposed algorithm in Figure \ref{fig:MainLP}. The effects of the  clamping operation are discussed in the next subsection.

\subsubsection{The clamping operation and its effects}
Suppose that $x^*_c$ is the result of the clamping operation on $x^*$. Note that there exists a $Z^*_c\in \mathcal{R}^{N\times |L|}$ such that $x^*_c = Z^*_c\times S$. As a result, note the following inequalities:
\begin{equation}
\begin{array}{l}
\displaystyle \hspace{2cm}||Ax^*_c-b||_{\infty} = \max_{k\in K} \left\{|f_k^T(x^*_c-x^*)|\right\} \leq ||D||_{\infty}\left(\frac{2r}{N}\right),
\end{array}
\label{eqn:bound1}
\end{equation}
and
\begin{equation}
\begin{array}{l}
\displaystyle\hspace{0.5cm} \left|c^T\mbox{vec}(Z^*_c) - c^T\mbox{vec}(Z^*)\right| = \frac{1}{N}\left| \textbf{1}^{1\times N}\times \left(Z^*_c-Z^*\right)\times S_p\right| \leq ||D_p||_{\infty}\left(\frac{2r}{N}\right),
\end{array}
\label{eqn:bound2}
\end{equation}
where
\begin{equation}
\begin{array}{l}
\displaystyle \hspace{0.25cm} D = \left[(L_2-L_1)~~(L_2-L_1)~\dots~(L_{m-1}-L_{m-2})~~(L_m-L_{m-1})\right]\\
\displaystyle D_p = \left[\left|L_2^2-L_1^2\right|~~\left|L_2^2-L_1^2\right|~\dots~\left|L_{m-1}^2-L_{m-2}^2\right|~~\left|L_m^2-L_{m-1}^2\right|\right].
\end{array}
\end{equation}
From the inequality in \eqref{eqn:bound1}, it is clear that the solution $x^*_c$ satisfies the harmonic constraints within a bound given by $\displaystyle ||D||_{\infty}\left(\frac{2r}{N}\right)$, which can be made arbitrarily small by increasing $N$. 

The focus is now to assess the impact clamping operation on the optimality of the total harmonic distortion. In the previous section, it was pointed out that minimizing the THD is equivalent to minimizing the total energy in the signal. Recall that the total energy for $x^*$ is given by $\displaystyle \frac{1}{N}\sum_{i=1}^N(x^*(i))^2$. Using Jensen's inequality (see \cite{bernstein2009matrix}) one obtains 
\begin{equation}
\begin{array}{l}
\displaystyle \frac{1}{N}\sum_{i=1}^N(x^*(i))^2 = \frac{(x^*)^Tx^*}{N} \leq c^T\mbox{vec}(Z^*) = \sum_{i=1}^N\left(Z^*(i,:)\times S_p\right).
\end{array}
\label{eqn:jensen}
\end{equation}
The extent of the difference $\displaystyle\left(c^T\mbox{vec}(Z^*)-\frac{(x^*)^Tx^*}{N}\right)$ can be obtained as a scalar multiple of the solution to a convex optimization problem. The magnitude of the difference is bounded by:
\begin{equation}
\begin{array}{l}
\displaystyle \hspace{0.5cm}\left(c^T\mbox{vec}(Z^*)-\frac{(x^*)^Tx^*}{N}\right) \leq \frac{(2r)}{N}\delta,
\end{array}
\label{eqn:jensenbound}
\end{equation}
where
\begin{equation}
\begin{array}{l}
\displaystyle \delta = \max_{\lambda\in \mathcal{R}^{m\times 1}} ~~\left(\lambda^TS_p - \left(\lambda^T S\right)^2\right)\\
\displaystyle \hspace*{0.75cm} \mbox{subject to}\\
\displaystyle \hspace*{1.25cm}\lambda \geq 0 \mbox{~~and~~} \lambda^T\textbf{1}^{m\times 1}=1.
\end{array}
\label{eqn:jensenbound1}
\end{equation}
Note that
\begin{equation}
\displaystyle \frac{1}{N}\left|(x_c^*)^Tx_c^*- (x^*)^Tx^*\right| = \left|c^T\mbox{vec}(Z^*_c) - \frac{(x^*)^Tx^*}{N}\right|.
\label{eqn:equality}
\end{equation}
Using triangle inequality, one obtains
\begin{equation}
\begin{array}{l}
\displaystyle \left|c^T\mbox{vec}(Z^*_c) - \frac{(x^*)^Tx^*}{N}\right| \leq \left|c^T\mbox{vec}(Z^*_c) - c^T\mbox{vec}(Z^*)\right| + \left|c^T\mbox{vec}(Z^*) - \frac{(x^*)^Tx^*}{N}\right|,\\
\displaystyle \hspace*{4cm} \leq ||D_p||_{\infty}\left(\frac{2r}{N}\right) + \frac{(2r)}{N}\delta.
\end{array}
\label{eqn:bound3}
\end{equation}
The inequality in \eqref{eqn:bound3}, together with \eqref{eqn:equality}, implies that the energy of the final output of the main algorithm shown in Figure \ref{fig:MainLP}, which is equal to $c^T\mbox{vec}(Z^*_c)$ is bounded within $\left(\displaystyle ||D_p||_{\infty}\left(\frac{2r}{N}\right) + \frac{(2r)}{N}\delta\right)$ range of the total energy of $x^*$ given by $\displaystyle \frac{(x^*)^Tx^*}{N}$. 
Reiterating, one has the following inequalities
\begin{equation}
\frac{(x^*)^Tx^*}{N} \leq c^T\mbox{vec}(Z^*) \leq c^T\mbox{vec}(Z^*_c),
\end{equation}
and, the difference between the first and the last term can be made arbitrarily small by choosing a sufficiently large $N$. Moreover, the center term represents the optimal value of the linear program in \eqref{eqn:MainLP}. Since THD is a continuous function of the total energy, the THD of $x^*_c$ obtained from the algorithm shown in Figure \ref{fig:MainLP} converges to the optimal value of \eqref{eqn:prob_definition_4} as $N\rightarrow \infty$.

\section{Simulations}
This section presents simulation results of the proposed algorithm in this paper for generating multilevel PWM with a prescribed set of harmonics and the lowest THD. Selective Harmonic Elimination and Harmonic Compensation present an interesting case for this part of the paper. Several scenarios are chosen: 3-level, 5-level, 8-level and 11-level PWMs, each for SHE and HC. 

\begin{figure*}[!ht]
    \centering
    \begin{subfigure}
        \centering
        \includegraphics[angle=0, scale=0.22]{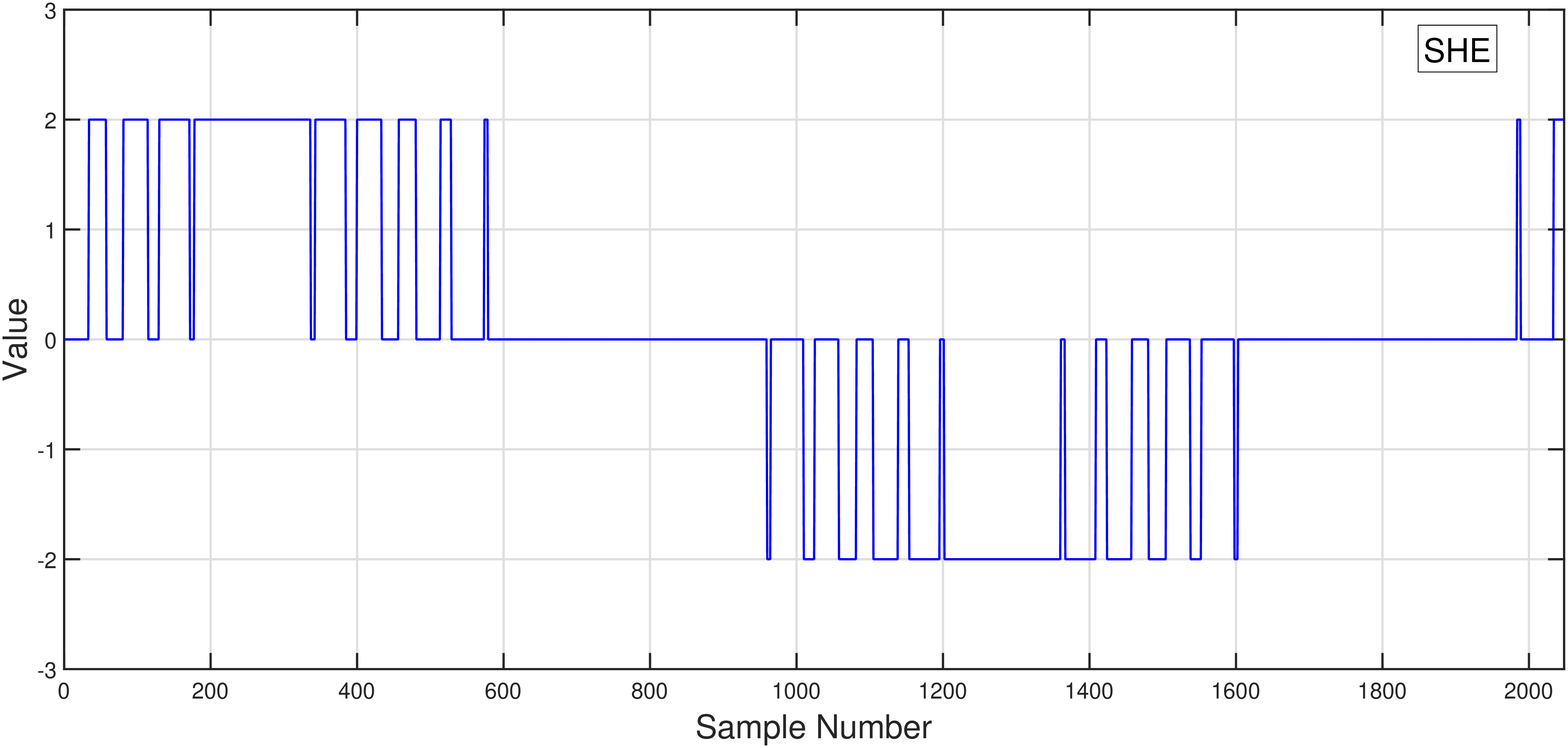}
    \end{subfigure}%
    \hspace{0.5cm} 
    \begin{subfigure}
        \centering
        \includegraphics[angle=0, scale=0.22]{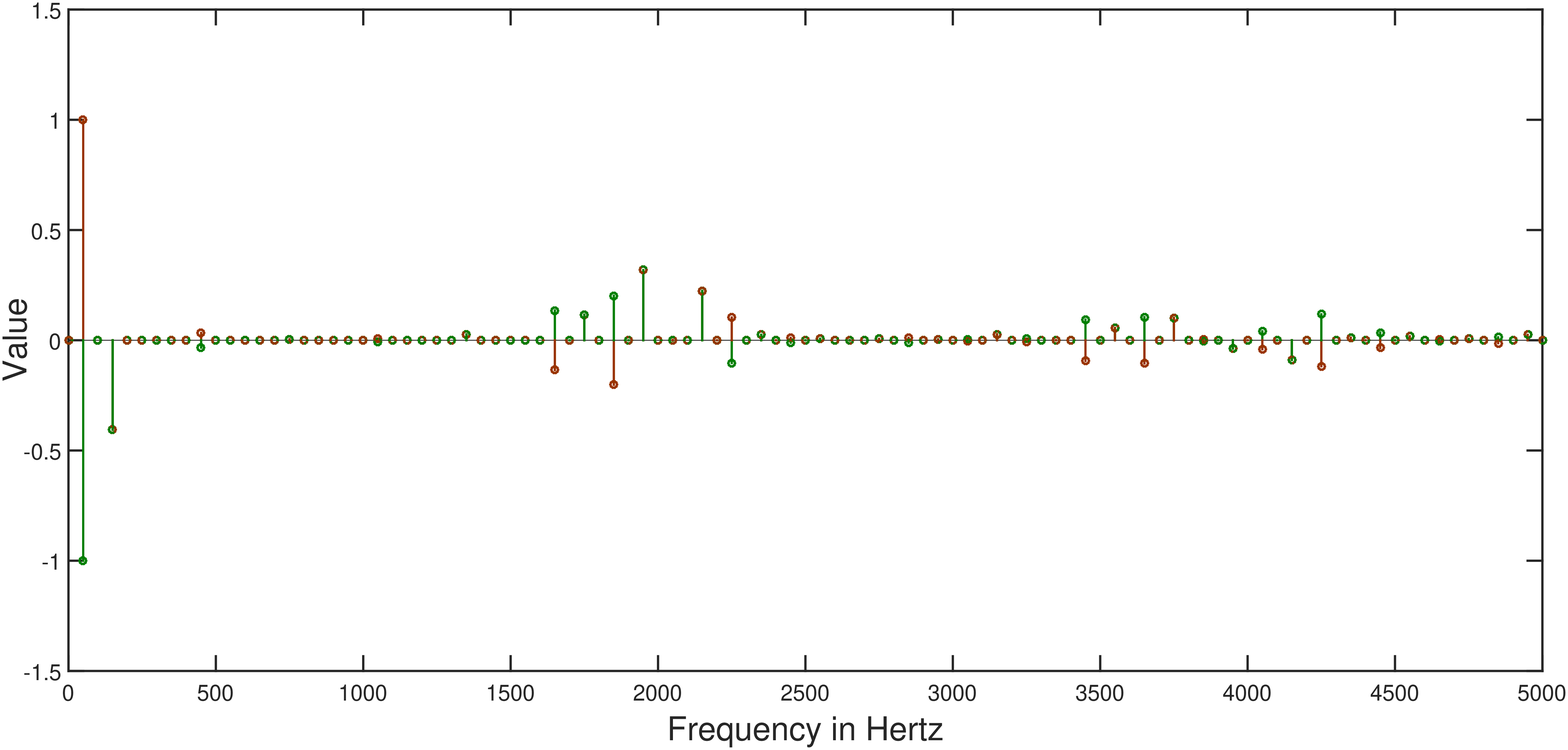}
    \end{subfigure}\\
    \begin{subfigure}
        \centering
        \includegraphics[angle=0, scale=0.22]{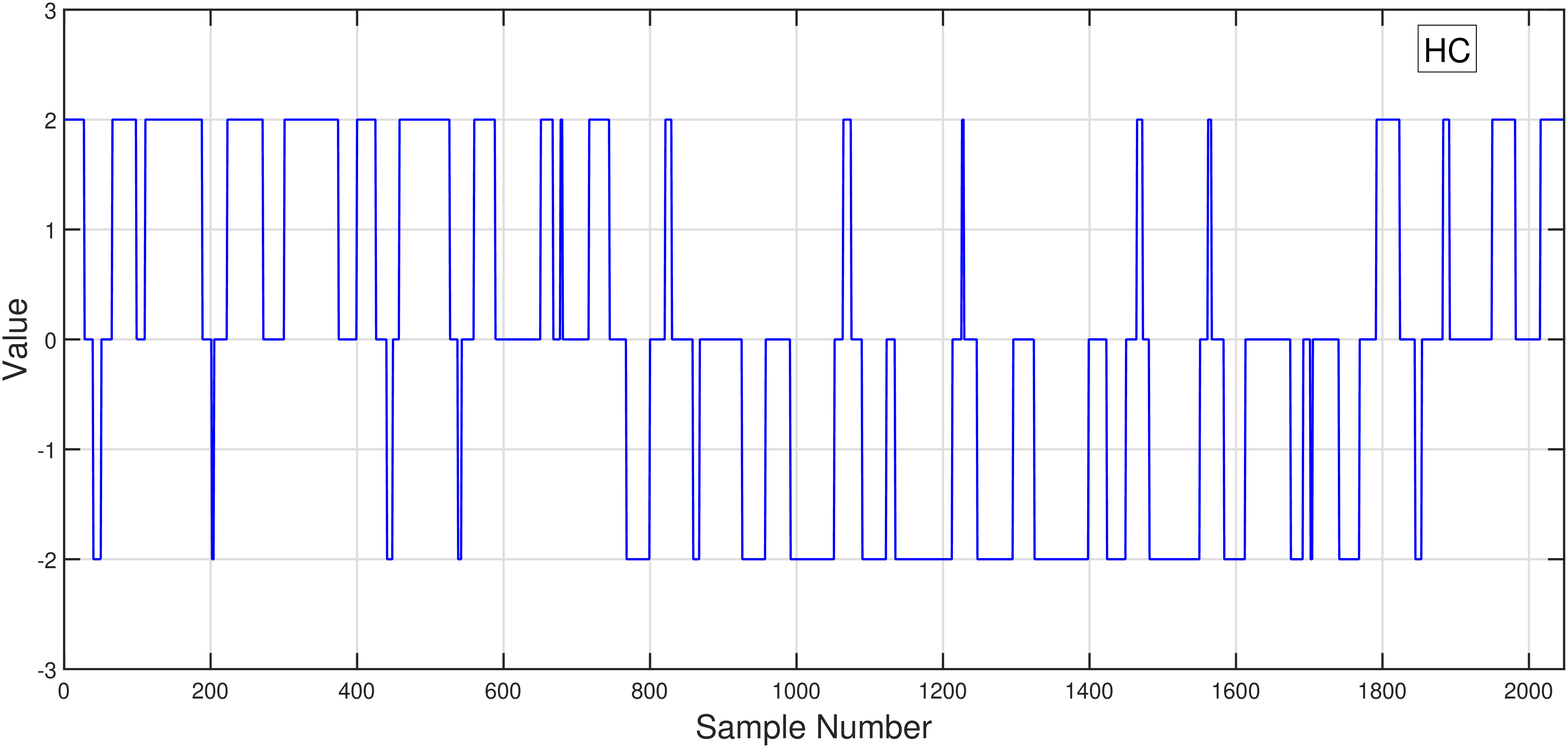}
    \end{subfigure}%
    \hspace{0.5cm}
    \begin{subfigure}
        \centering
        \includegraphics[angle=0, scale=0.22]{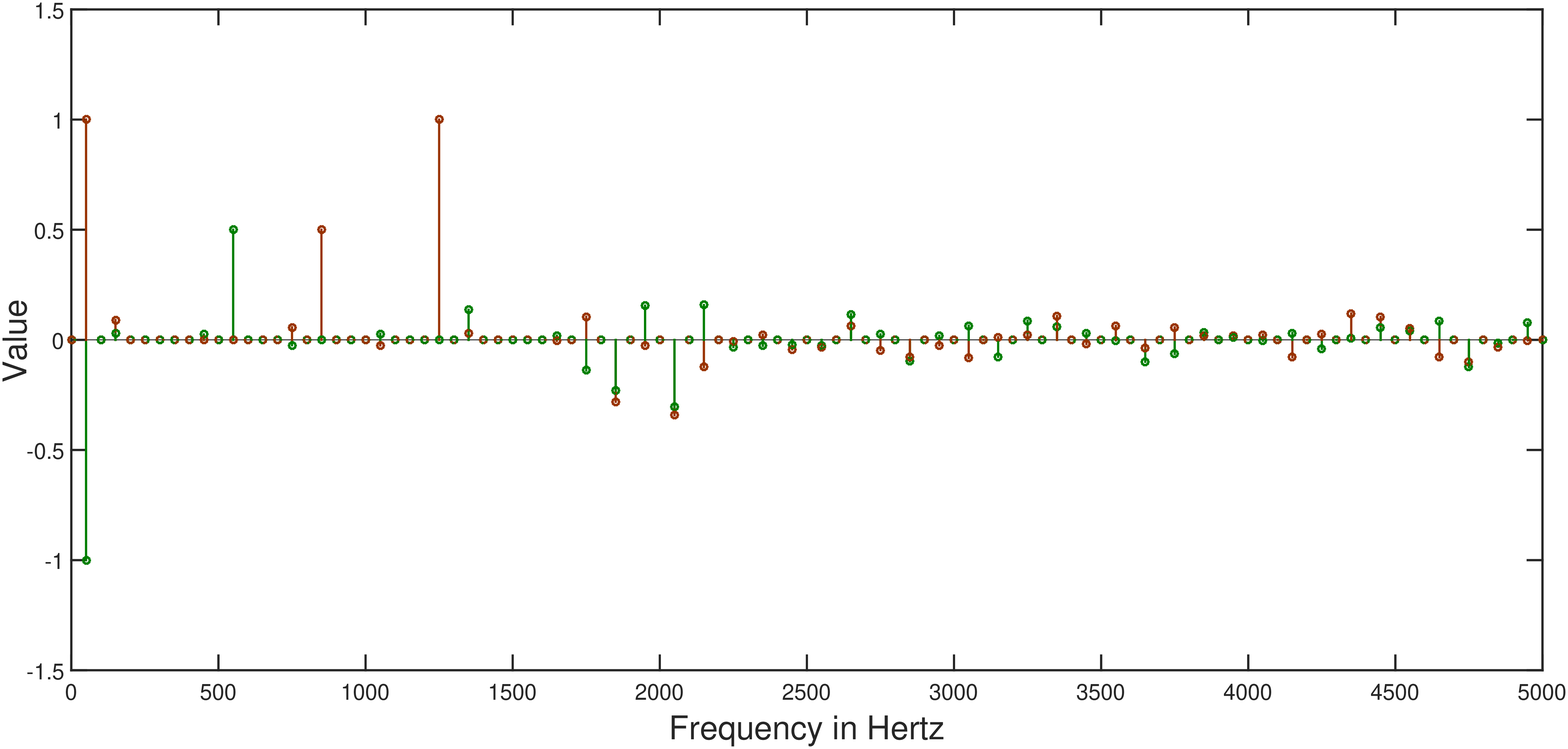}
    \end{subfigure}
    
    \caption{\textbf{Design of 3-level PWM}: This figure shows the PWMs obtained from the  main algorithm shown in Figure \ref{fig:MainLP} when the number of samples $N = 2048$, the set of levels $S = [-2~~~ 0~~~ 2]$, the prescribed harmonic numbers are given by $K = [1~~~ 5~~~ 7~~~ 11~~~ 13~~~ 17~~~ 19~~~ 23~~~ 25~~~ 29~~~ 31]$, and the respective harmonic quantities are given by $h = [1-1i~~~ 0~~~ 0~~~ 0~~~ 0~~~ 0~~~ 0~~~ 0~~~ 0~~~ 0~~~ 0]$ for \textbf{SHE} and $h = [1-1i~~~ 0~~~ 0~~~ 0.5i~~~ 0~~~ 0.5~~~ 0~~~ 0~~~ 1~~~ 0~~~ 0]$ for \textbf{HC}. Both the waveforms are accompanied by their respective Fourier series. The red and green coloured stems represent the cosine and the sine components present in the corresponding waveforms The THD of the output waveforms are 0.3601 and 0.2215, respectively. See Tables \ref{table:SHE}-\ref{table:HC} for more details.} 
    \label{fig:level3}
\end{figure*}

\begin{figure*}[!ht]
    \centering
    \begin{subfigure}
        \centering
        \includegraphics[angle=0, scale=0.22]{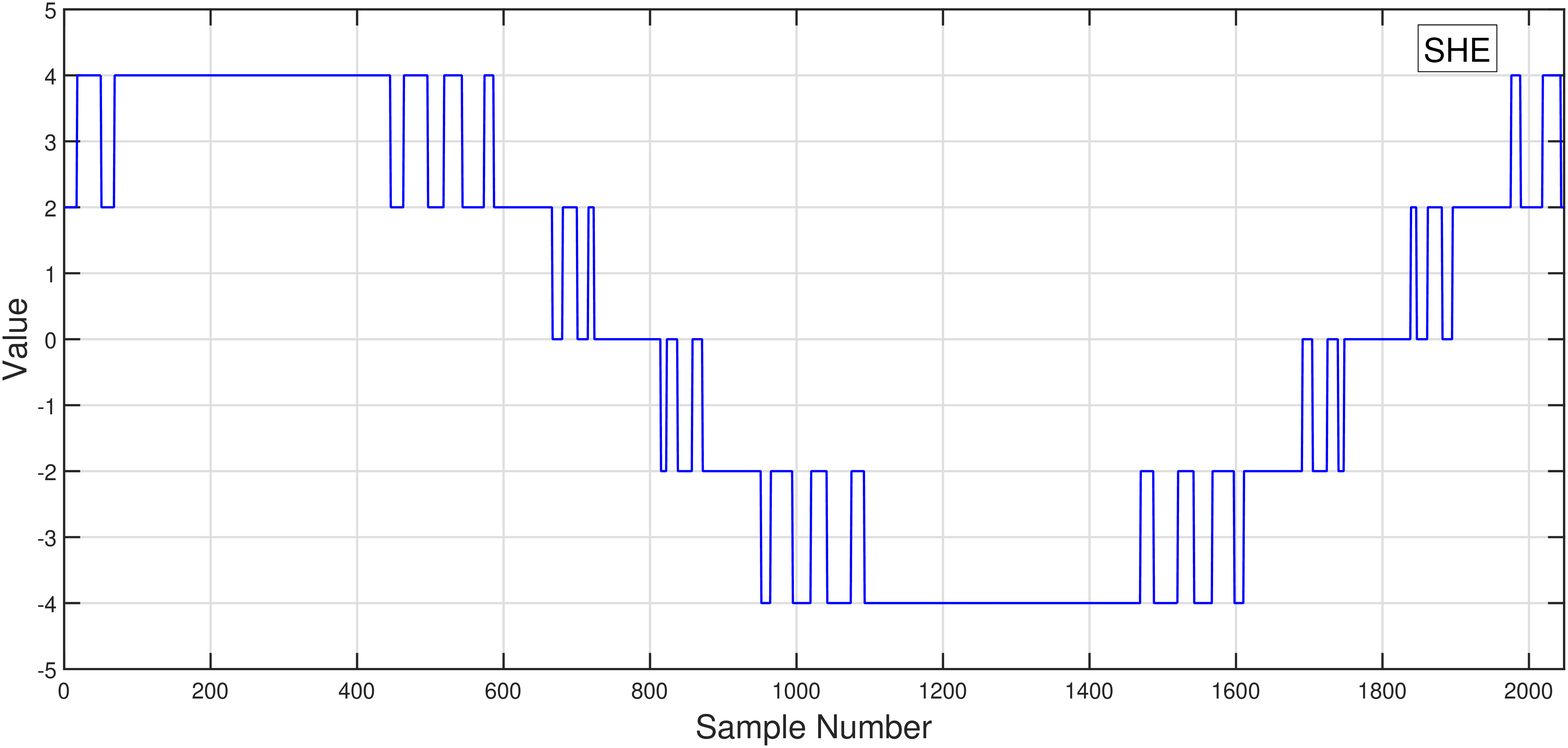}
    \end{subfigure}%
    \hspace{0.5cm} 
    \begin{subfigure}
        \centering
        \includegraphics[angle=0, scale=0.22]{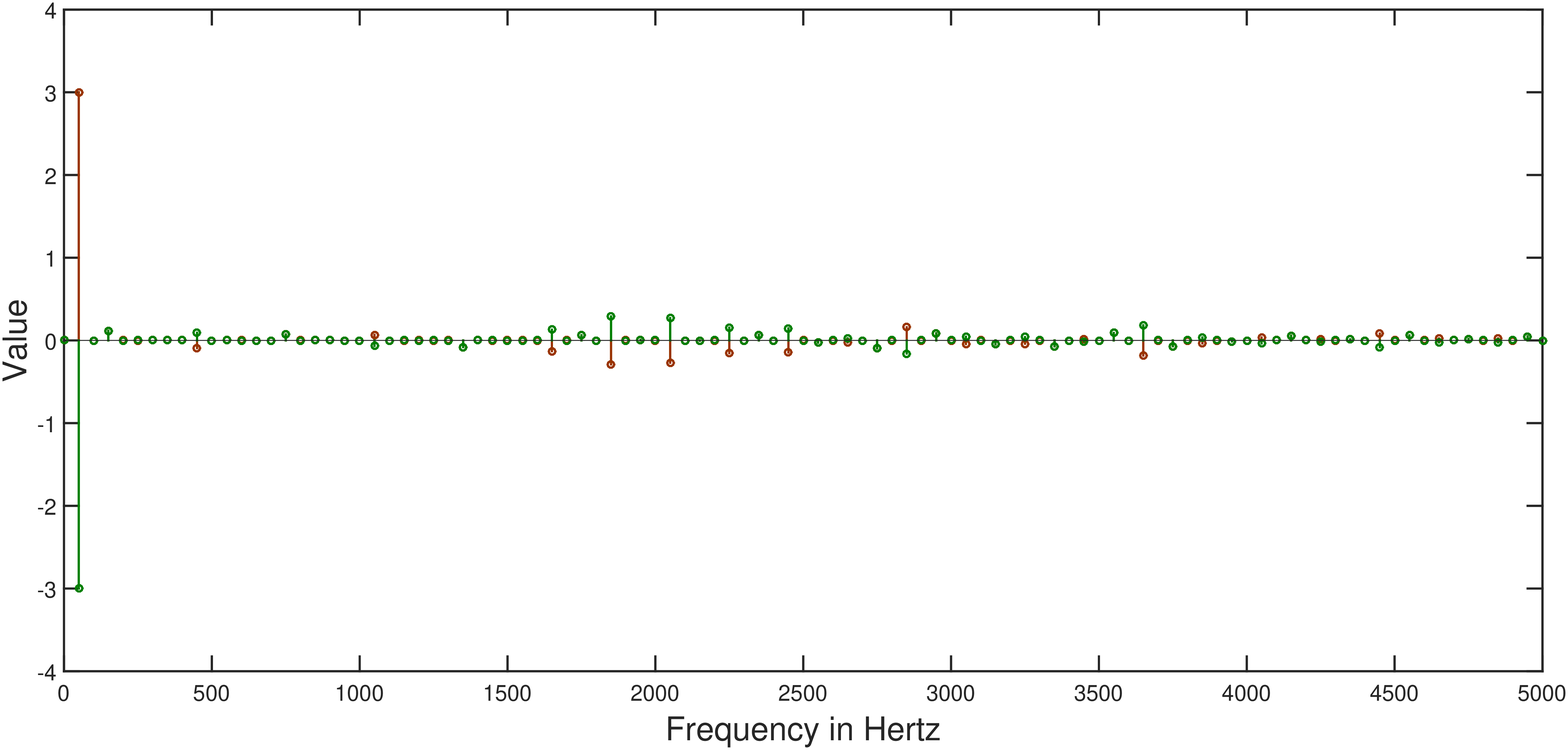}
    \end{subfigure}\\
    \begin{subfigure}
        \centering
        \includegraphics[angle=0, scale=0.22]{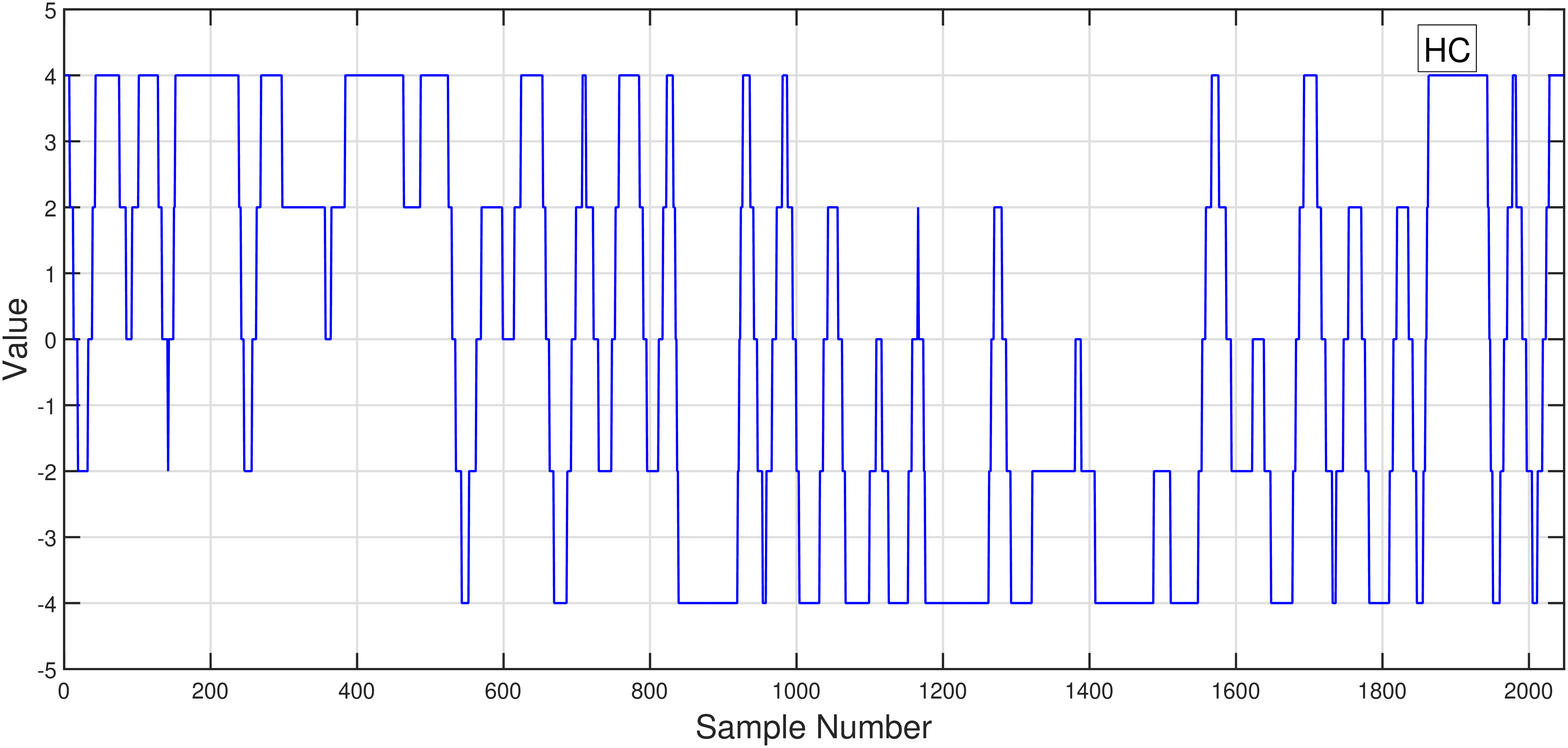}
    \end{subfigure}%
    \hspace{0.5cm}
    \begin{subfigure}
        \centering
        \includegraphics[angle=0, scale=0.22]{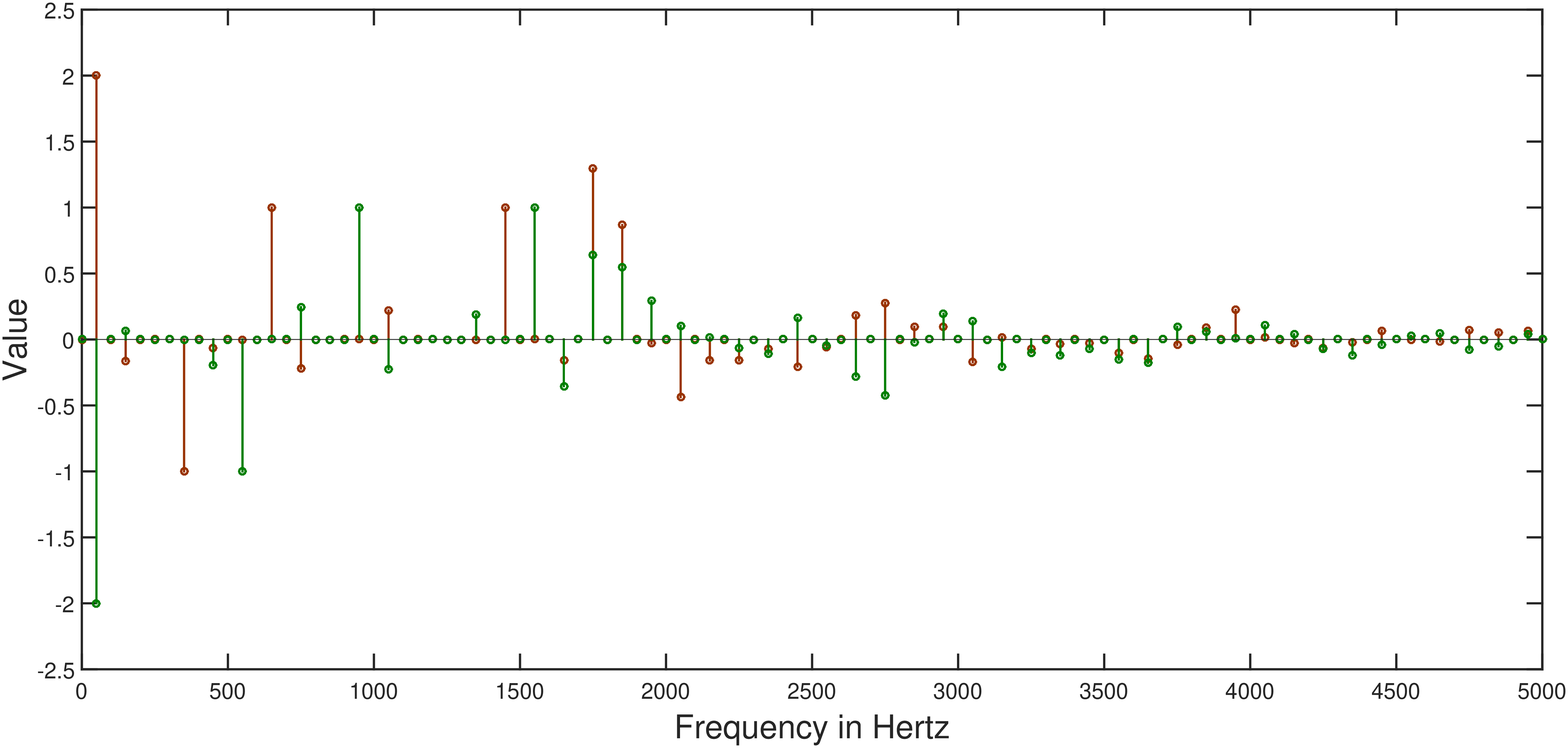}
    \end{subfigure}
    
    \caption{\textbf{Design of 5-level PWM}: This figure shows the PWMs obtained from the  main algorithm shown in Figure \ref{fig:MainLP} when the number of samples $N = 2048$, the set of levels $S = [-4~~~ -2~~~ 0~~~ 2~~~ 4]$, the prescribed harmonic numbers are given by $K = [1~~~ 5~~~ 7~~~ 11~~~ 13~~~ 17~~~ 19~~~ 23~~~ 25~~~ 29~~~ 31]$, and the respective harmonic quantities are given by $h = [3-3i~~~ 0~~~ 0~~~ 0~~~ 0~~~ 0~~~ 0~~~ 0~~~ 0~~~ 0~~~ 0]$ for \textbf{SHE} and $h = [2-2i~~~ 0~~~ -1~~~ -1i~~~ 1~~~ 0~~~ 1i~~~ 0~~~ 0~~~ 1~~~ 1i]$ for \textbf{HC}. Both the waveforms are accompanied by their respective Fourier series. The red and green coloured stems represent the cosine and the sine components present in the corresponding waveforms The THD of the output waveforms are 0.0511 and 0.2726, respectively. See Tables \ref{table:SHE}-\ref{table:HC} for more details.}
    \label{fig:level5}
\end{figure*}

\begin{figure*}[!ht]
    \centering
    \begin{subfigure}
        \centering
        \includegraphics[angle=0, scale=0.22]{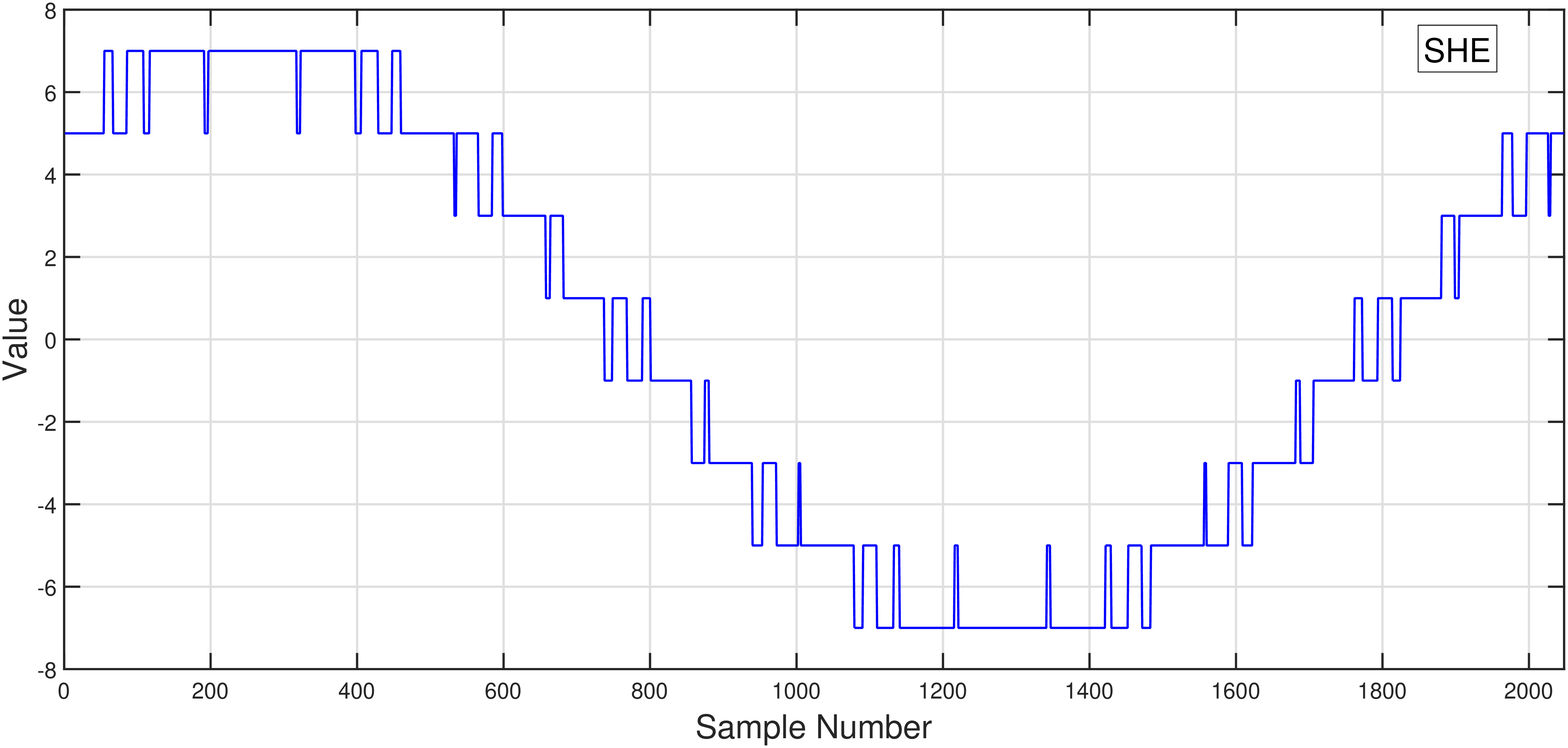}
    \end{subfigure}%
    \hspace{0.5cm} 
    \begin{subfigure}
        \centering
        \includegraphics[angle=0, scale=0.22]{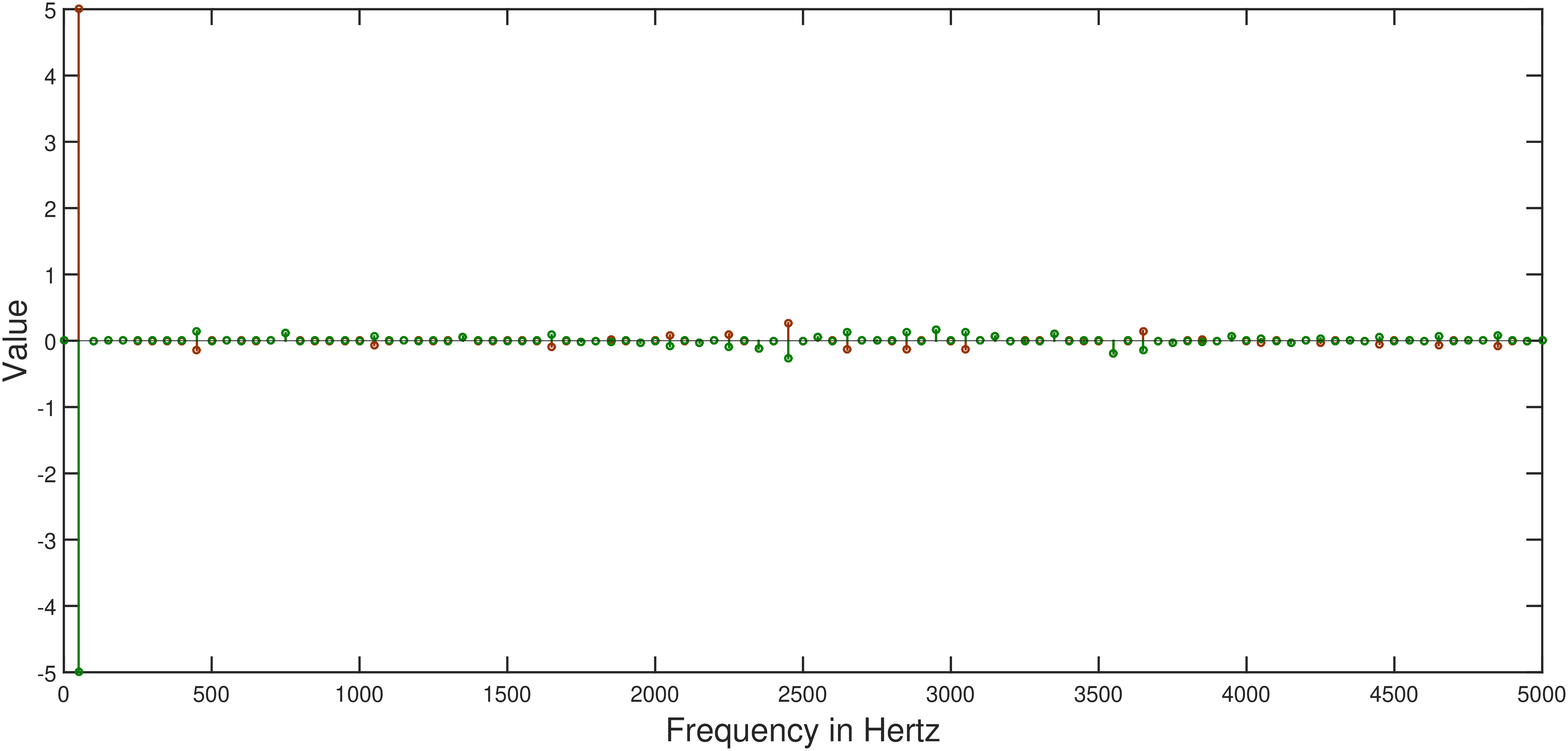}
    \end{subfigure}\\
    \begin{subfigure}
        \centering
        \includegraphics[angle=0, scale=0.22]{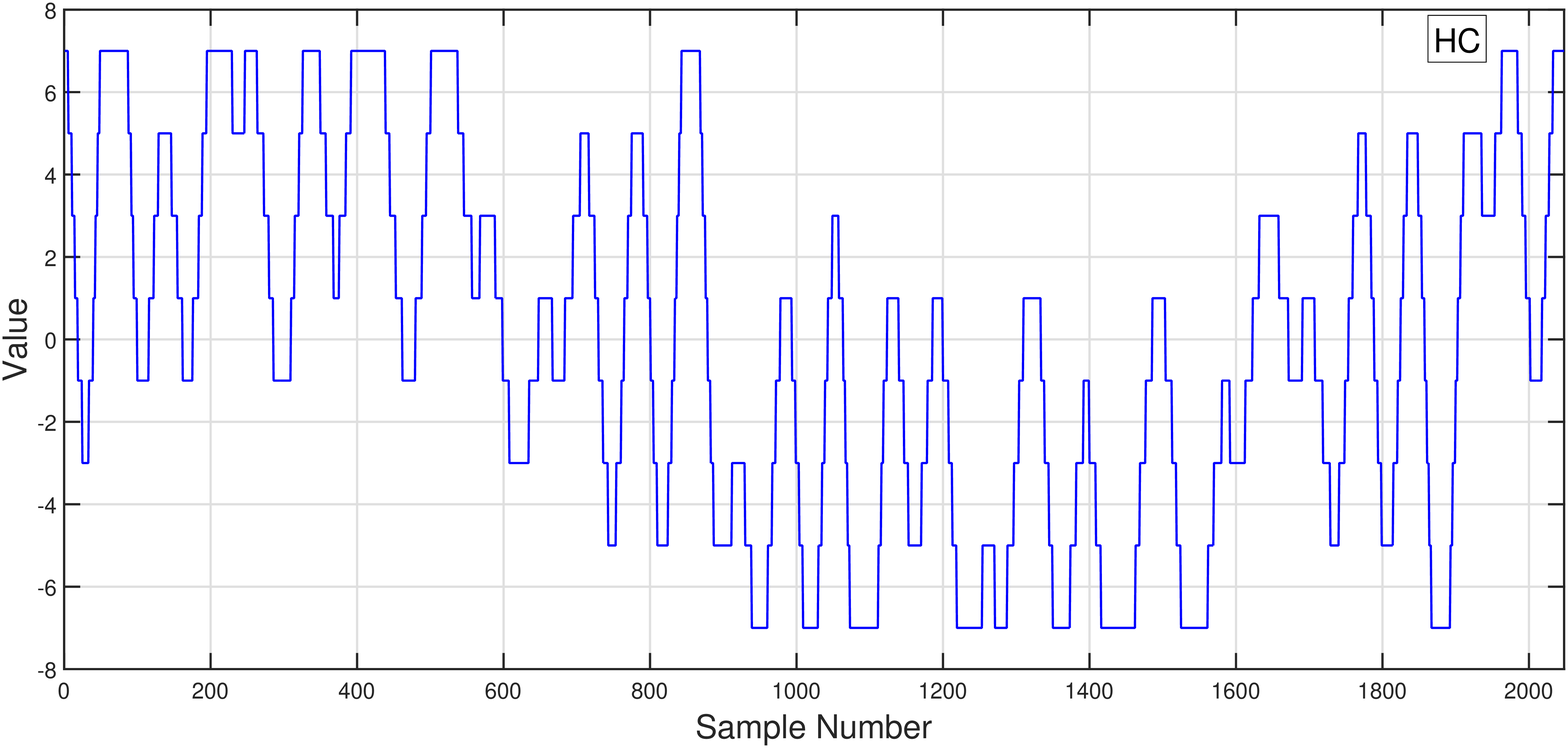}
    \end{subfigure}%
    \hspace{0.5cm}
    \begin{subfigure}
        \centering
        \includegraphics[angle=0, scale=0.22]{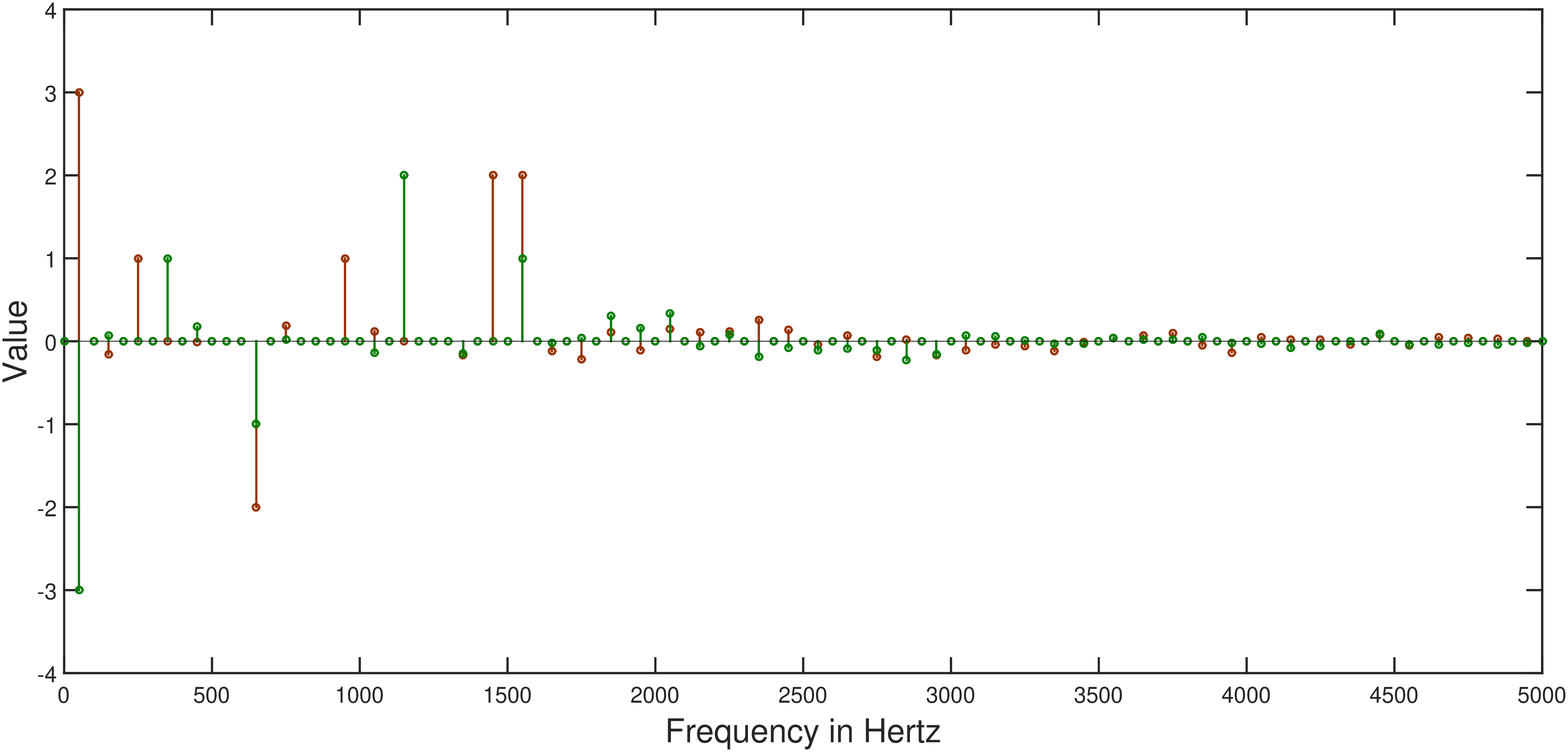}
    \end{subfigure}
    
    \caption{\textbf{Design of 8-level PWM}: This figure shows the PWMs obtained from the  main algorithm shown in Figure \ref{fig:MainLP} when the number of samples $N = 2048$, the set of levels $S = [-7~~~ -5~~~ -3~~~ -1~~~ 1~~~ 3~~~ 5~~~ 7~~~]$, the prescribed harmonic numbers are given by $K = [1~~~ 5~~~ 7~~~ 11~~~ 13~~~ 17~~~ 19~~~ 23~~~ 25~~~ 29~~~ 31]$, and the respective harmonic quantities are given by $h = [5-5i~~~ 0~~~ 0~~~ 0~~~ 0~~~ 0~~~ 0~~~ 0~~~ 0~~~ 0~~~ 0]$ for \textbf{SHE} and $h = [3-3i~~~ 1~~~ 1i~~~ 0~~~ -2-1i~~~ 0~~~ 1~~~ 2i~~~ 0~~~ 2~~~ 2+1i]$ for \textbf{HC}. Both the waveforms are accompanied by their respective Fourier series. The red and green coloured stems represent the cosine and the sine components present in the corresponding waveforms The THD of the output waveforms are 0.0191 and 0.0362, respectively. See Tables \ref{table:SHE}-\ref{table:HC} for more details.}
    \label{fig:level8}
\end{figure*}

\begin{figure*}[!ht]
    \centering
    \begin{subfigure}
        \centering
        \includegraphics[angle=0, scale=0.22]{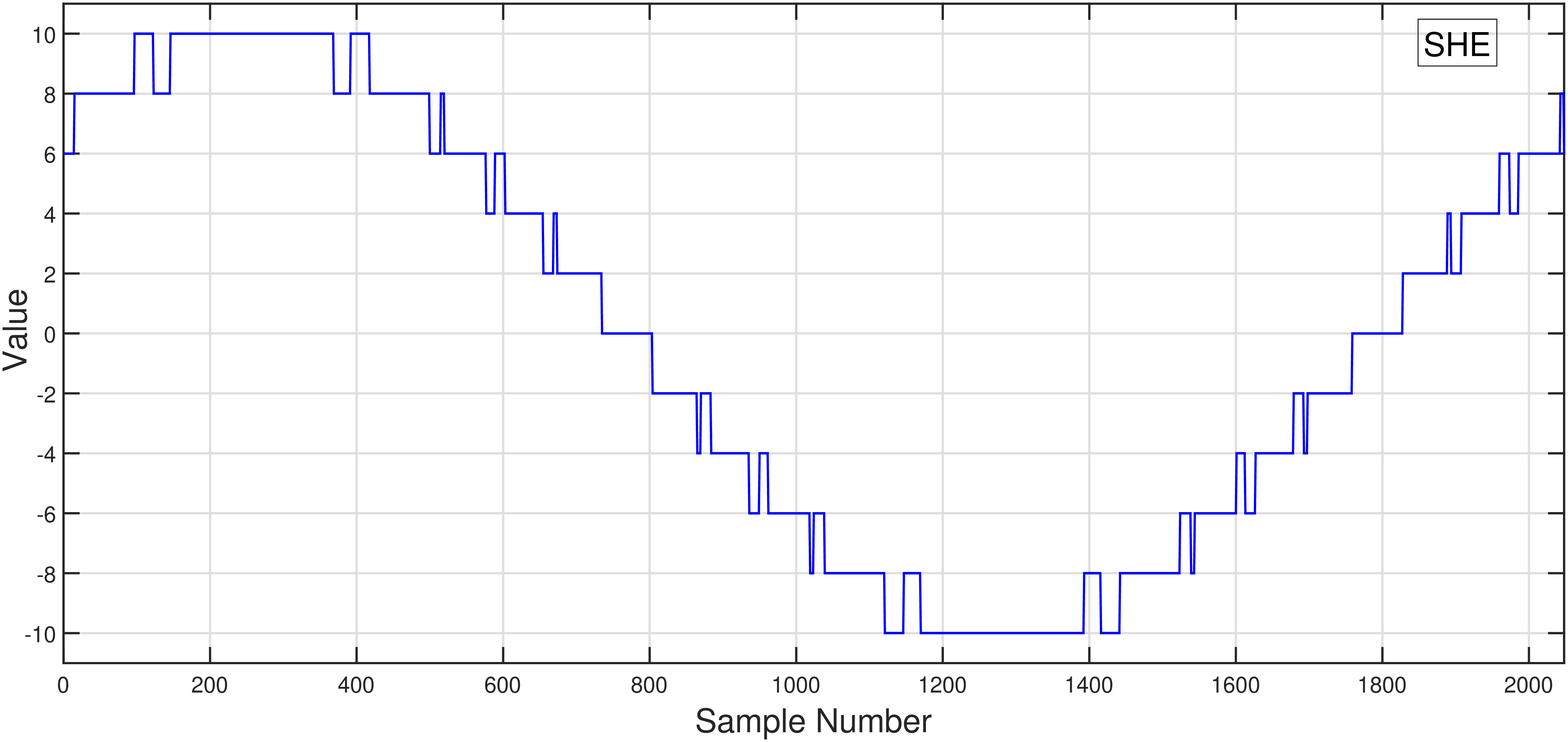}
    \end{subfigure}%
    \hspace{0.5cm} 
    \begin{subfigure}
        \centering
        \includegraphics[angle=0, scale=0.22]{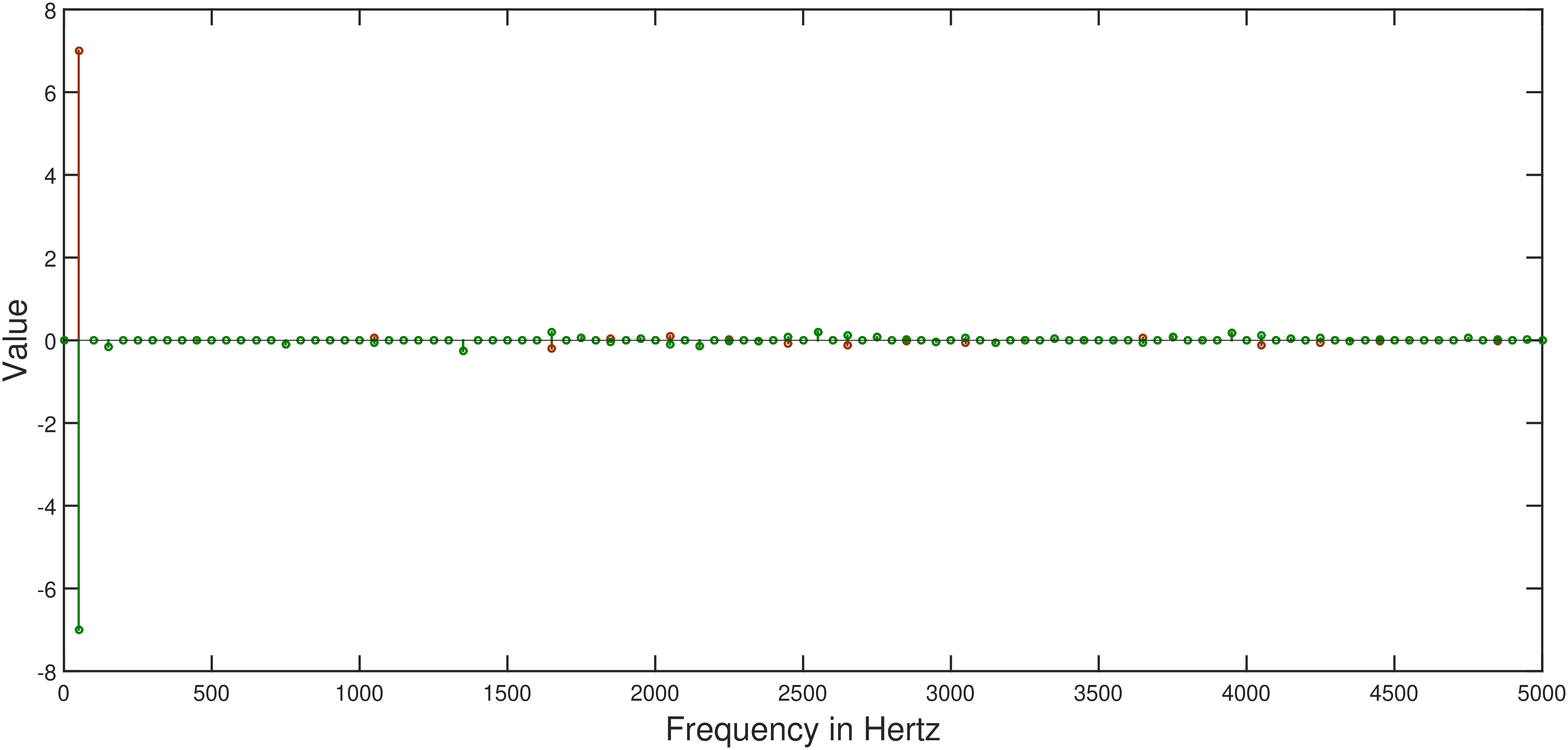}
    \end{subfigure}\\
    \begin{subfigure}
        \centering
        \includegraphics[angle=0, scale=0.22]{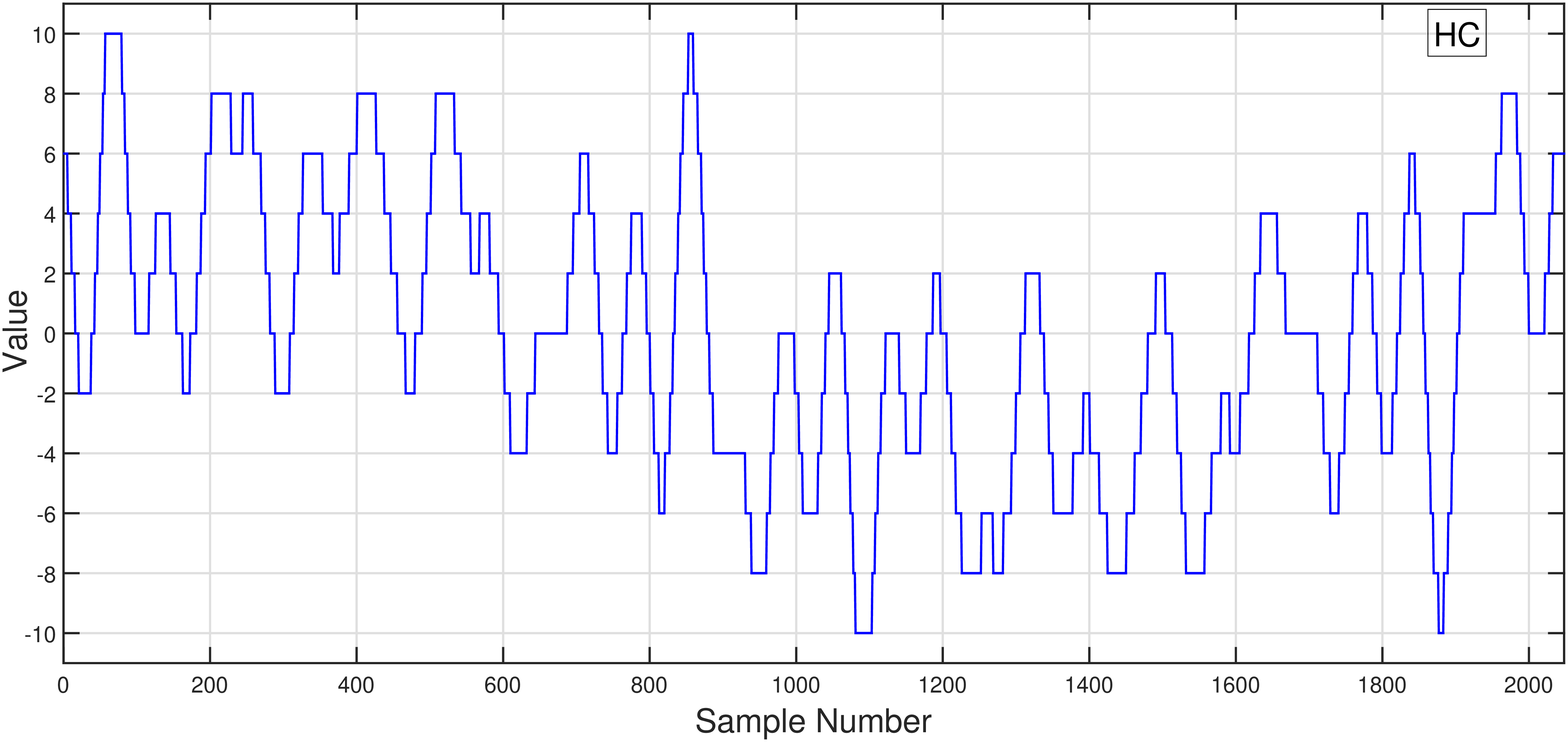}
    \end{subfigure}%
    \hspace{0.5cm}
    \begin{subfigure}
        \centering
        \includegraphics[angle=0, scale=0.22]{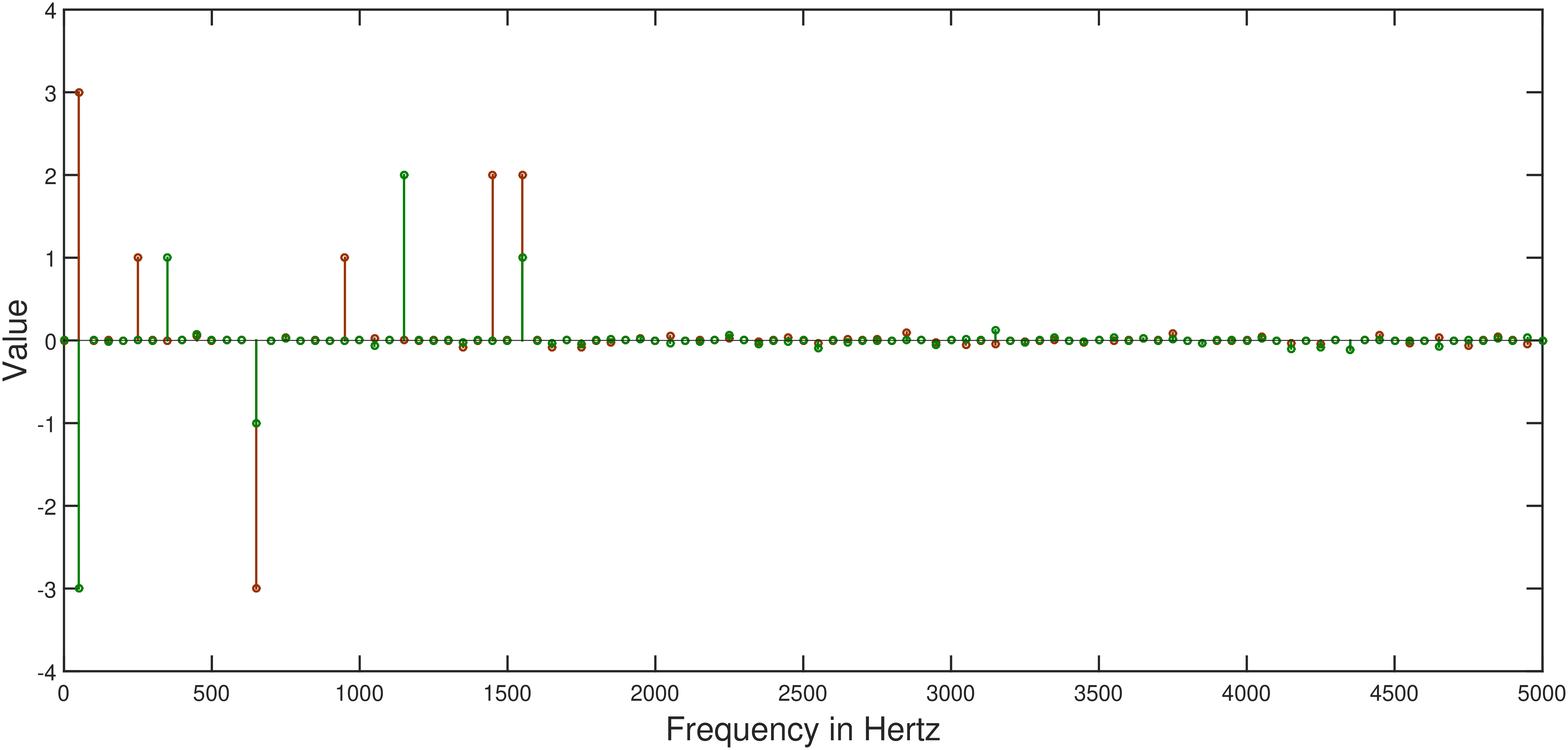}
    \end{subfigure}
    
    \caption{\textbf{Design of 11-level PWM}: This figure shows the PWMs obtained from the  main algorithm shown in Figure \ref{fig:MainLP} when the number of samples $N = 2048$, the set of levels $S = [-10~~~ -8~~~ -6~~~ -4~~~ -2~~~ 0~~~ 2~~~ 4~~~ 6~~~ 8~~~ 10]$, the prescribed harmonic numbers are given by $K = [1~~~ 5~~~ 7~~~ 11~~~ 13~~~ 17~~~ 19~~~ 23~~~ 25~~~ 29~~~ 31]$, and the respective harmonic quantities are given by $h = [7-7i~~~ 0~~~ 0~~~ 0~~~ 0~~~ 0~~~ 0~~~ 0~~~ 0~~~ 0~~~ 0]$ for \textbf{SHE} and $h = [3-3i~~~ 1~~~ 1i~~~ 0~~~ -3-1i~~~ 0~~~ 1~~~ 2i~~~ 0~~~ 2~~~ 2+1i]$ for \textbf{HC}. Both the waveforms are accompanied by their respective Fourier series. The red and green coloured stems represent the cosine and the sine components present in the corresponding waveforms The THD of the output waveforms are 0.0090 and 0.0272, respectively. See Tables \ref{table:SHE}-\ref{table:HC} for more details.}
    \label{fig:level11}
\end{figure*}

The simulations were done on MATLAB using CVX (\cite{grant2008cvx}, \cite{boyd2004convex}). The details of the prescribed harmonics and their values for the illustrations are given in Table \ref{table:SHE} and Table \ref{table:HC}. In particular for SHE, the cosine and sine Fourier components corresponding to $\displaystyle 5^{\rm{th}}, 7^{\rm{th}}, 11^{\rm{th}}, 13^{\rm{th}}, 17^{\rm{th}}, 19^{\rm{th}}, 23^{\rm{rd}}, 25^{\rm{th}}, 29^{\rm{th}}, 31^{\rm{st}}\}$ harmonics are set to zero while the fundamental cosine and sine harmonics are set to non-zero values. For all the simulations, $\displaystyle N=2048$, in accordance with $\displaystyle (r\ll N)$ and $\displaystyle (\max(K)\ll N)$. In addition, the average value of the PWM is constrained to be zero for all the simulations. The output waveforms are shown in Figures \ref{fig:level3}-\ref{fig:level11}. Each waveform represents one time period. The waveforms are accompanied by their Discrete Fourier Transform (DFT) which shows their harmonic content.     

It might be necessary for some engineering applications to generate waveforms with zero even harmonics. This translates to the a half-wave anti-symmetry in the output, that is, $x(k)=-x(N-k),~~\forall 1\leq k\leq N/2, ~k\in \mathcal{Z}$. This in turn boils down to rewriting the constraints and the cost functions in the variables given by the reduced set $\displaystyle \left\{x\left(i\right)| i\in \mathcal{Z} \mbox{~and~} 1\leq i\leq N/2\right\}$. The qualitative aspects of the proposed algorithm and the solution remain the same as discussed earlier. In these simulations, the levels chosen are symmetric around 0, but an asymmetric level set can also be considered for suitable applications. 
\begin{table}[th!]
\centering
\caption{This table lists the simulation scenarios for \textbf{SHE} used in this paper. The set of levels, the prescribed harmonic numbers and their harmonic values are tabulated. The multilevel PWMs are generated using the proposed algorithm shown in Figure \ref{fig:MainLP}. The waveforms are depicted in Figure \ref{fig:level3}-\ref{fig:level11}. The total harmonic distortion of the obtained PWMs are also listed in the last column of the table.}
\label{table:SHE}
\begin{tabular}{|c|c|c|c|}
\hline
Levels ($L$)                                                                               & \begin{tabular}[c]{@{}c@{}}Prescribed\\ Harmonics ($K$)\end{tabular}                                            & \begin{tabular}[c]{@{}c@{}}Harmonic\\ Values ($H$)\end{tabular}                                                          & THD    \\ \hline
\{-2, 0, 2\}                                                                         & \multirow{4}{*}{\begin{tabular}[c]{@{}c@{}}\{1, 5, 7, 11, 13, 17,\\ 19, 23, 25, 29, 31\}\end{tabular}} & \begin{tabular}[c]{@{}c@{}}$h_c$ = {[}1 0 0 0 0 0 0 0 0 0 0{]}\\ $h_s$ = {[}-1 0 0 0 0 0 0 0 0 0 0{]}\end{tabular} & 0.3601 \\ \cline{1-1} \cline{3-4} 
\{-4, -2, 0, 2, 4\}                                                                  &                                                                                                        & \begin{tabular}[c]{@{}c@{}}$h_c$ = {[}3 0 0 0 0 0 0 0 0 0 0{]}\\ $h_s$ = {[}-3 0 0 0 0 0 0 0 0 0 0{]}\end{tabular} & 0.0511 \\ \cline{1-1} \cline{3-4} 
\{-7, -5, -3, -1, 1, 3, 5, 7\}                                                       &                                                                                                        & \begin{tabular}[c]{@{}c@{}}$h_c$ = {[}5 0 0 0 0 0 0 0 0 0 0{]}\\ $h_s$ = {[}-5 0 0 0 0 0 0 0 0 0 0{]}\end{tabular} & 0.0191 \\ \cline{1-1} \cline{3-4} 
\begin{tabular}[c]{@{}c@{}}\{-10, -8, -6, -4, -2,\\ 0, 2, 4, 6, 8, 10\}\end{tabular} &                                                                                                        & \begin{tabular}[c]{@{}c@{}}$h_c$ = {[}7 0 0 0 0 0 0 0 0 0 0{]}\\ $h_s$ = {[}-7 0 0 0 0 0 0 0 0 0 0{]}\end{tabular} & 0.0090 \\ \hline
\end{tabular}
\end{table}

\begin{table}[th!]
\centering
\caption{This table lists the simulation scenarios for \textbf{HC} used in this paper. The set of levels, the prescribed harmonic numbers and their harmonic values are tabulated. The multilevel PWMs are generated using the proposed algorithm shown in Figure \ref{fig:MainLP}. The waveforms are depicted in Figure \ref{fig:level3}-\ref{fig:level11}. The total harmonic distortion of the obtained PWMs are also listed in the last column of the table.}
\label{table:HC}
\begin{tabular}{|c|c|c|c|}
\hline
Levels ($L$)                                                                              & \begin{tabular}[c]{@{}c@{}}Desired\\ Harmonics ($K$)\end{tabular}                                            & \begin{tabular}[c]{@{}c@{}}Harmonic\\ Values ($H$)\end{tabular}                                                                  & THD    \\ \hline
\{-2, 0, 2\}                                                                         & \multirow{4}{*}{\begin{tabular}[c]{@{}c@{}}\{1, 5, 7, 11, 13, 17,\\ 19, 23, 25, 29, 31\}\end{tabular}} & \begin{tabular}[c]{@{}c@{}}$h_c$ = {[}1 0 0 0 0 0.5 0 0 1 0 0{]}\\ $h_s$ = {[}-1 0 0 0.5 0 0 0 0 0 0 0{]}\end{tabular} & 0.2215 \\ \cline{1-1} \cline{3-4} 
\{-4, -2, 0, 2, 4\}                                                                  &                                                                                                        & \begin{tabular}[c]{@{}c@{}}$h_c$ = {[}2 0 -1 0 1 0 0 0 0 1 0{]}\\ $h_s$ = {[}-2 0 0 -1 0 0 1 0 0 0 1{]}\end{tabular}       & 0.2726 \\ \cline{1-1} \cline{3-4} 
\{-7, -5, -3, -1, 1, 3, 5, 7\}                                                       &                                                                                                        & \begin{tabular}[c]{@{}c@{}}$h_c$ = {[}3 1 0 0 -2 0 1 0 0 2 2{]}\\ $h_s$ = {[}-3 0 1 0 -1 0 0 2 0 0 1{]}\end{tabular}       & 0.0362 \\ \cline{1-1} \cline{3-4} 
\begin{tabular}[c]{@{}c@{}}\{-10, -8, -6, -4, -2,\\ 0, 2, 4, 6, 8, 10\}\end{tabular} &                                                                                                        & \begin{tabular}[c]{@{}c@{}}$h_c$ = {[}3 1 0 0 -3 0 1 0 0 2 2{]}\\ $h_s$ = {[}-3 0 1 0 -1 0 0 2 0 0 1{]}\end{tabular}       & 0.0272 \\ \hline
\end{tabular}
\end{table}


\section{Conclusion}
The problem of designing a multilevel switched periodic waveform with a prescribed set of harmonics was considered. It was formulated as a linear program by discretizing the integrals defining the harmonic constraints. The cost function was chosen with the aim of minimizing the total harmonic distortion. The solution to the linear program was shown to have a small fraction of values not in the level set. These values were clamped to the nearest values in the level set. The clamping operation ensured the multilevel characteristic of the solution. It was shown that the clamped solution satisfied the harmonic constraints within arbitrarily small bounds. In addition, the total harmonic distortion also converged to its optimal value with finer discretization. This approach was applied to obtain waveforms for various cases of Selective Harmonic Elimination and Harmonic Compensation, and their total harmonic distortion was reported.

As for future directions of work, it would interesting to explore if a similar method can be developed for designing multilevel PWMs with a given finite autocorrelation sequence. Another avenue of research can be the development of multilevel sigma-delta modulators which produce minimal harmonic distortion.

\end{document}